\documentclass[10pt,superscriptaddress,twocolumn,aps,showpacs,nofootinbib,eqsecnum,twoside,prc]{revtex4}
\usepackage[pdfborder={0 0 0},colorlinks=true,linkcolor=blue,citecolor=blue,filecolor=blue,urlcolor=blue]{hyperref}   
\usepackage{url}
\usepackage{ulem}
\usepackage{slashed}
\usepackage{amssymb}
\usepackage{amsfonts}
\usepackage{mathbbol} 
\usepackage{amstext}
\usepackage{graphicx}
\usepackage{color}
\usepackage{epsfig,amsmath,lscape}
\usepackage{soul}
\usepackage{mathptmx}                
\usepackage{dcolumn}                 
\usepackage{bm}                      
\usepackage[caption=false]{subfig}
\usepackage{subfig}

\begin{document}

\title{Liquid-gas phase transition in strange hadronic matter with relativistic models}

\author{James R. Torres}
\affiliation{Departamento de F\'{\i}sica - CFM - Universidade Federal de Santa Catarina, \\ 
Florian\'opolis - SC - CP. 476 - CEP 88.040 - 900 - Brazil \\ email: james.r.torres@posgrad.ufsc.br}

\author{F. Gulminelli}
\affiliation{CNRS and ENSICAEN, UMR6534, LPC, \\ 
14050 Caen c\'edex, France \\ email:
gulminelli@lpccaen.in2p3.fr }

\author{D\'ebora P. Menezes}
\affiliation{Departamento de F\'{\i}sica - CFM - Universidade Federal de Santa Catarina, \\ 
Florian\'opolis - SC - CP. 476 - CEP 88.040 - 900 - Brazil \\ email: debora.p.m@ufsc.br}

\begin{abstract}
\begin{description}
\item[Background] 
The advent of new dedicated experimental programs on hyperon physics 
is rapidly boosting the field, and the possibility of synthetizing multiple strange 
hypernuclei requires the addition of the strangeness degree of freedom to 
the models dedicated to nuclear structure and nuclear matter studies 
at low energy. 
\item[Purpose] 
We want to settle the influence of strangeness on the nuclear liquid-gas phase transition. Because of the large uncertainties concerning the hyperon sector, we do not aim at a quantitative estimation of the phase diagram but rather at a qualitative description of the phenomenology, as model independent as possible.  
\item[Method]
We analyze the phase diagram of low density matter composed of neutrons, protons and $\Lambda$ hyperons using a Relativistic Mean Field (RMF) model. We largely explore the parameter space to pin down generic features of the phase transition, and compare the results to ab-initio quantum Monte Carlo calculations.
\item[Results] 
We show that the liquid-gas phase transition is only slightly quenched by the addition of hyperons. Strangeness is seen to be an order parameter of the phase transition, meaning that dilute strange matter is expected to be unstable with respect to the formation of hyper-clusters. 
\item[Conclusions] 
More quantitative results within the RMF model need improved functionals
at low density, possibly fitted to ab-initio calculations of nuclear and $\Lambda$ matter.
\end{description}
\end{abstract}


\maketitle


\section{Introduction}

\paragraph*{}
It is well known that nuclear matter below saturation exhibits a first-order phase transition belonging to the Liquid-Gas (LG) universality class \cite{Finn82,Bertsch83,Muller95,Glendenning01,Chomaz04,Das05,
Ducoin06,Rios10,Wellenhofer15}. The study of the associated phase diagram is not only a playground for many-body theorists, but it is also of clear relevance for nuclear phenomenology, since the very existence of atomic nuclei can be understood as a finite size manifestation of that phase transition. 
In a similar way, one can ask whether the existence of hypernuclei as bound systems implies the presence of a similar phase transition in the extended phase diagram where strangeness represents an extra dimension.
\paragraph*{}
Since the first synthesis of $\Lambda$-hypernuclei in the 80's, numerous nuclear matter studies including hyperons have been performed \cite{Millener88,Mares89,Rufa90,Schaffner94,Schaffner02}. 
These early studies assumed very attractive couplings in the strange sector in order to justify the extra binding measurements of double $\Lambda$-hypernuclei \cite{Aoki91,Dover91}. 
As a consequence, it was predicted that multi-strange clusters and even strangelets could be stable and  possibly accessible in heavy-ion collisions. In particular in Ref.\cite{Schaffner02}, 
the occurrence of a thermodynamic phase transition in strange compressed baryonic matter was predicted,  which would lead to a new family of neutron stars characterized by much smaller 
radii than usually considered. 
\paragraph*{}
However, more recent analysis \cite{Aoki09,Ahn13} of double $\Lambda$-hypernuclei tend to suggest a very small attraction in the $\Lambda-\Lambda$ channel, 
and the stability of pure $\Lambda$-matter seems to be ruled out. Most hypernuclear matter studies are nowadays essentially motivated by assessing the 
strange content of neutron star cores, and therefore concentrate on matter in $\beta$-equilibrium \cite{Schaffner08}. At $\beta$-equilibrium, no hyperons 
appear below baryonic densities of the order of 3$\rho_0$ or more. For this reason, the influence of   strangeness  on the low density nuclear matter phase 
diagram was never studied to our knowledge. Still, the existence of single and double $\Lambda$-hypernuclei, and the very active research on multiply strange 
nuclei with the advent of new dedicated experimental programs such as J-Parc in Japan or PANDA at FAIR \cite{Minato13,Ikram14,Sang14, Khan15} suggests that the
nuclear liquid-gas phase transition should be preserved by the consideration of the strangeness degree of freedom \cite{Mallik15}.
\paragraph*{}
In this paper, we explore the influence of strangeness on the LG phase transition with popular Relativistic Mean-Field (RMF) models. Like in any other phenomenological effective model, the couplings of the RMF are not fully known even at subsaturation densities. In particular,  neutron star physics has taught us in the recent years
that it is important to go beyond a simple SU(6)
or even SU(3) symmetry, and extra attractive $\sigma^*$ and repulsive $\phi$ mesons specifically coupled to the strange baryons should be introduced \cite{Sym04,Weiss,Sym05,Luiz,Gusakov,Shen06}, which leads to a potentially uncontrolled multiplication of parameters.
However, if we limit ourselves to the simple system composed of neutrons, protons and $\Lambda$-hyperons, nuclear and hypernuclear structure provide some constraints that can be used to limit the parameter space of the model. 
In this paper, we consider the simple linear and non-linear Walecka model for the np$\Lambda$ system, and discuss the modification of the nuclear matter phase diagram under a wide variation of  coupling constants, in the acceptable parameter space constrained  both from hypernuclear data and ab-initio calculations of hypernuclear matter.
We show that in the whole parameter space the LG phase transition is preserved by the addition of strangeness, even if the extension of the spinodal along the strange density direction is subject to large uncertainties. The instability zone is globally quenched by strangeness, but the strange density is an order parameter of the transition. This means that from the thermodynamic point of view, the formation of hyperclusters with multiple  $\Lambda$'s should  be favored at low density \cite{Ikram14,Sang14,Khan15}, which has possible implications in relativistic heavy ion collisions \cite{Mallik15}.
\paragraph*{}
The paper is organized as follows: section \ref{Formalism} shortly recalls the main equations of the Walecka model, both in its linear and non-linear version, 
for the np$\Lambda$ system with inclusion of strange mesons. Section \ref{couplings} defines the 
coupling parameter space of the model, under the constraints of well defined values for the $\Lambda$-potential as requested by the available hypernuclear data. 
To further refine the domain of acceptable parameters, Section \ref{abinitio} compares the RMF functionals with recent ab-initio predictions of n$\Lambda$-matter 
with the Auxiliary Field Diffusion Monte Carlo (AFDMC) technique \cite{abinitio00}.
In Section \ref{spinodal} the general formalism for the analysis of spinodal instabilities in multi-component systems is revisited. 
The main results of our work are presented in Section \ref{results}, which shows in detail the instability properties 
of n$\Lambda$ and np$\Lambda$ matter with the different choices for the couplings. Finally Section \ref{summary} summarizes the paper. 

\section{Formalism} \label{Formalism}

\paragraph*{}
In this section, we present the hadronic Equation Of State (EOS) used in this work. 
We describe matter within the framework of  Relativistic Mean Field (RMF) models involving the interaction of Dirac baryons mediated by the scalar and vector mesons which are independent degrees of freedom \cite{Original RMF,Walecka,GlennBook,Glenn00,Glenn01,Boguta,Kapusta00,Kapusta01}. 
The  scalar-isoscalar $\sigma$ field mediates the medium-range attraction between baryons, the vector-isoscalar $\omega$ field mediates the short-range repulsion between baryons, 
the strange scalar $\sigma^{\ast}$ field mediates the medium-range attraction between hyperons, strange vector $\phi$ field mediates the short-range repulsion between 
hyperons and finally the $\rho$ meson field allows us to adjust isovector properties of nuclear matter. In the present work, we used the Nonlinear Walecka Model (NLWM) 
and the Linear Walecka Model (LWM), which can be obtained by just turning off the nonlinear terms, in the presence of the mesons listed above.
Nonlinear means that there are also self interaction terms for the scalar field $\sigma$ in the Lagrangian density, as proposed by Boguta and Bodmer \cite{Boguta,Boguta01}, 
what provides better results than the LWM \cite{GlennBook}. The Lagrangian density reads:

\begin{eqnarray*}
\mathcal{L} &\mathcal{=}&\sum_{j}\overline{\psi }_{j}\left[ \gamma^{\mu }\left( i\partial _{\mu }-g_{\omega j}\omega _{\mu }-g_{\phi j}\phi_{\mu } \right.\right. \nonumber \\
&& \left.\left.-g_{\rho j}\vec{\tau}\cdot \vec{\rho}_{\mu }\right) -m_{j}^{\ast}\right] \psi _{j}\nonumber \\
\end{eqnarray*}%

\begin{eqnarray}
&&+\frac{1}{2}\left( \partial _{\mu }\sigma \partial ^{\mu }\sigma-m_{\sigma }^{2}\sigma ^{2}\right) \nonumber \\
&&-\frac{1}{3}bM_{N}\left(g_{\sigma N}\sigma\right)^{3} -\frac{1}{4}c\left(g_{\sigma N}\sigma\right)^{4} \nonumber \\   
&&+\frac{1}{2}\left( \partial _{\mu }\sigma^{\ast}\partial^{\mu }\sigma^{\ast}-m_{\sigma }^{2} \sigma^{\ast 2} \right) \nonumber \\
&&-\frac{1}{4} \Omega_{\mu \nu} \, \Omega^{\mu \nu}+ \frac{1}{2} m_{\omega}^2 \, \omega_\mu \omega^\mu \nonumber \\
&&-\frac{1}{4}\Phi _{\mu \nu }\Phi ^{\mu \nu }+\frac{1}{2}m_{\phi}^{2}\phi ^{\mu }\phi _{\mu }, \nonumber \\
&&-\frac{1}{4} \vec R_{\mu \nu } \,\cdot\,  \vec R^{\mu \nu } \nonumber \\
&&+ \frac{1}{2} m_\rho^2 \, \vec \rho_{\mu} \,\cdot\, \vec \rho^{\, \mu},
\label{lagrangian}
\end{eqnarray}%

where $m_j^* = m_j - g_{\sigma j}\, \sigma- g_{ \sigma^{\ast} j}\, \sigma^{\ast}$ is the baryon effective mass and $m_j$ is the bare mass of the baryon $j$.
The terms $\Omega_{\mu\nu }=\partial_\mu \omega_\nu - \partial_\nu \omega_\mu$~, $\Phi_{\mu\nu }=\partial_\mu \phi_\nu - \partial_\nu \phi_\mu$ and
$\vec R_{\mu \nu } = \partial_\mu \vec \rho_\nu -\partial_\nu \vec \rho_\mu 
-g_{\rho j} \left({\vec \rho_\mu \,\times \, \vec \rho_\nu } \right)$ are the strength tensors, where the up arrow in the last term denotes the isospin vectorial space with the $\vec \tau$  isospin operator.
The coupling constants are $g_{ij}=\chi_{ij} g_{i N}$, with the mesons denoted by index $i = \sigma, \omega, \rho, \sigma^{\ast}, \phi$ and the baryons denoted by $j$. Note that $\chi_{ij}$ is a proportionality factor between $g_{ij}$ and the nucleon coupling constants $g_{i N}$, with $N=n,~p$. 
The couplings $b$ and $c$ are the weights of the nonlinear scalar terms. The sum over $j$ can be extended over all baryons of the octet
$\left(n~, p~, \Lambda~,\Sigma^-~,\Sigma^0~,\Sigma^+~,\Xi^-~,\Xi^0~ \right)$.

\paragraph*{}
The values ​​of the coupling constants of the nucleons with mesons $\sigma$, $\omega$ and $\rho$ are obtained from the phenomenology. These constants are tuned to the bulk properties of nuclear matter. 
Some of these properties are not known exactly, just within certain ranges, like the effective masses of the nucleons, therefore there are many sets of parameters that describe the bulk properties.
The biggest uncertainties concern the hyperon coupling constants, because the   phenomenological information from hypernuclei is not sufficient to completely pin down the interaction in the strange sector \cite{Hyper03_ULN,
Hyper02_ULN}. The hyperon couplings are chosen in different ways in the literature, either 
based on simple symmetry considerations \cite{Sym00,Moszkowski,Pal,Sym01,Sym02,Luiz},
or requiring an EOS in $\beta$-equilibrium sufficiently stiff to justify the observation of very massive neutron stars \cite{Glenn02,Glenn03}. 

\paragraph*{}
Some different approaches, all affected by a certain degree of arbitrariness, are listed here: 1) Some authors argue that $\chi_{\sigma j}=\chi_{\omega j}=\chi_{\rho j}=\sqrt{2/3}$ \cite{Moszkowski}; 2) In another work \cite{Pal}, the authors claim that $\chi_{\sigma \Lambda}=\chi_{\omega\Lambda}=\chi_{\sigma \Sigma}=\chi_{\omega \Sigma}=2/3$ , 
$\chi_{\sigma \Xi}=\chi_{\omega \Xi}=1/3$, $\chi_{\rho \Lambda}=0$, $\chi_{\rho \Sigma}=2$ and $\chi_{\rho \Xi}=0$; 3) Based on the experimental analysis of $\Lambda$-hypernuclei data, an alternative constraint is given by
$U_{\Lambda}\left(n_{N}=n_{0}\right)=\chi_{\omega
  \Lambda}\left(g_{\omega N}\right)-\chi_{\sigma
  \Lambda}\left(g_{\sigma N}\right)=-28~\text{MeV}$ for the fixed
$\chi_{\sigma \Lambda}=0.75$. This last case can be extended to the
whole baryonic octet, indexed 
by $j$, setting $\chi_{\sigma j}=0.75$, $\chi_{\omega j}$ is given by
the above constraint and $\chi_{\rho h}=0$, where $h$ is the hyperon
index \cite{Glenn02}; 4) Taking into account the resulting  neutron star maximum mass \cite{Glenn02,Glenn03}.

\paragraph*{}
In the case of the inclusion of the strange mesons, $\sigma^{\ast}$ and $\phi$  \cite{Sym04,Weiss,Sym05}, we have to ensure that the nuclear matter properties are preserved when these new mesons are included. 
New mesons mean new interactions and also new constants, therefore the
arbitrariness introduced by these constants must be eliminated by data
whenever possible. In analogy with what has been done with the $g_{\sigma \Lambda}$, when constrained by the 
hypernuclear potential $U_{\Lambda}^{N}$ via hypernuclear data
\cite{Glenn02}, we can try to tie the strange constants 
to the $U_{\Lambda}^{\Lambda}$ data available in literature
\cite{Aoki09,Ahn13,Hyper00,Hyper01,Hyper02,Hyper03,Hyper00_ULN,
Hyper01_ULN,Hyper03_ULL,Hyper04_ULL,Hyper05_ULL}. In 
the next section we develop these ideas in detail.

\paragraph*{}
Applying the Euler-Lagrange equations to the lagrangian density Eq.(\ref{lagrangian}) and using the mean-field approximation \cite{Walecka}, 
($\sigma \to \langle \sigma \rangle = \sigma_{0}~ ;\;
\omega_\mu \to \langle \omega_\mu  \rangle = \delta_{\mu 0} \,\omega_0 ~;\; 
\vec \rho_\mu  \to \langle \vec \rho_\mu  \rangle = \delta_{\mu 0} \, \delta^{i 3} \rho_0^3  
\equiv \delta_{\mu 0} \, \delta^{i 3} \rho_{03} ~;\;  \sigma^{\ast} \to \langle \sigma^{\ast} \rangle = \sigma^{\ast}_{0}~ ;
\; \phi_\mu \to \langle \phi_\mu  \rangle = \delta_{\mu 0} \,\phi_0   $), we obtain the
following equations of motion for the meson fields at zero temperature:
\begin{equation*}
\left( g_{\sigma N}\sigma _{0}\right) =\Delta _{\sigma }\left( \sum_{j}\chi _{\sigma
j}\rho _{j}^{s}-bM_{n}\left( g_{\sigma N}\sigma _{0}\right) ^{2}-c\left(
g_{\sigma N}\sigma _{0}\right) ^{3}\right),
\end{equation*}%
\begin{equation*}
\left( g_{\omega N}\omega _{0}\right) =\Delta _{\omega }\sum_{j}\chi _{\omega j}n _{j},
\end{equation*}%
\begin{equation*}
\left( g_{\rho N}\rho _{0}\right) =\Delta _{\rho }\sum_{j}\tau _{3j}\chi _{\rho j}n _{j},
\end{equation*}%
\begin{equation*}
\left( g_{\sigma N}\sigma^{\ast}_{0}\right) =\Delta _{\sigma \sigma^{\ast} }\sum_{j}\chi _{\sigma^{\ast}j}\rho _{j}^{s},
\end{equation*}%
\begin{equation}
\left( g_{\omega N}\phi _{0}\right) =\Delta _{\omega \phi }\sum_{j}\chi _{\phi j}n _{j},
\end{equation}
where for simplicity we define the following factors:
$\Delta _{\sigma } =\left( \frac{g_{\sigma N}}{m_{\sigma }}\right) ^{2}$, 
$\Delta _{\omega } =\left( \frac{g_{\omega N}}{m_{\omega }}\right) ^{2}$,
$\Delta _{\rho } =\left( \frac{g_{\rho N}}{m_{\rho }}\right) ^{2}$,
$\Delta _{\sigma\sigma^{\ast}} =\left( \frac{g_{\sigma N}}{m_{\sigma^{\ast}}}\right) ^{2}$, 
$\Delta _{\omega \phi } =\left( \frac{g_{\omega N}}{m_{\phi }}\right) ^{2}$, $\chi _{\sigma j}~$,
$\chi _{\sigma^{\ast}j}~$, $\chi _{\omega j}~$, $\chi _{\rho j}$ and $\chi _{\phi j}$ are ratios between coupling constants and  $\tau _{3j}$ 
is the third component of the isospin projection of the $j$-baryon. The scalar and baryon densities are given respectively by
\begin{equation}
\rho _{j}^{s}=\frac{\gamma }{2\pi ^{2}}\int_{0}^{k_{Fj} }\frac{m_{j}^{\ast }}{\sqrt{p^{2}+m_{j}^{\ast2 }}}p^{2} dp
\end{equation}
and
\begin{equation}
n _{j}=\frac{\gamma }{2\pi ^{2}}\int_{0}^{k_{Fj} }p^{2}dp.
\end{equation}

\paragraph*{}
The energy density of the baryons is given by
\begin{equation}
\varepsilon _{B}=\frac{\gamma }{2\pi ^{2}}\sum%
\limits_{j}\int_{0}^{k_{Fj} }p^{2}\sqrt{p^{2}+m _{j}^{\ast2 }}dp
\label{baryon_energy_density}
\end{equation}

and for the mesons

\begin{eqnarray}
\varepsilon _{M}&=&\frac{\left( g_{\sigma N}\sigma _{0}\right) ^{2}}{%
2\Delta _{\sigma }}+\frac{\left( g_{\omega N}\omega
_{0}\right) ^{2}}{2\Delta _{\omega }}+\frac{\left( g_{\rho N}\rho
_{0}\right) ^{2}}{2\Delta _{\rho }} \nonumber \\ 
&&+\frac{\left( g_{\sigma N}\sigma^{\ast}_{0}\right)
^{2}}{2\Delta _{\sigma \sigma^{\ast}}}+\frac{\left( g_{\omega N}\phi
_{0}\right) ^{2}}{2\Delta _{\omega \phi }} \nonumber \\
&&+\frac{1}{3}bM_{n}\left( g_{\sigma N}\sigma _{0}\right) ^{3}+\frac{1}{4}c\left( g_{\sigma N}\sigma _{0}\right)^{4}.
\label{meson_energy_density}
\end{eqnarray}

Finally the total energy density is the summation

\begin{equation*}
\varepsilon =\varepsilon _{B}+\varepsilon _{M}.
\end{equation*}

\paragraph*{}
To obtain the chemical potential, one has to take the derivatives of the energy density with respect to the baryon density \cite{GlennBook}. 
Note the dependence of the Fermi momenta and the fields with the baryon density
in the upper limit of the integrals in Eq.(\ref{baryon_energy_density}) and Eq.(\ref{meson_energy_density}) respectively. Using the derivative 
chain rule and the equation of motion for the $\sigma$ field, we obtain
\begin{equation}
\mu _{j}^{\ast }=\mu _{j}-\chi _{\sigma j}\left( g_{\omega N}\omega _{0}\right) -\tau _{3j}\chi _{\rho j}\left( g_{\rho N}\rho _{0}\right) -\chi _{\omega j}\left( g_{\phi N}\phi _{0}\right).
\end{equation}

The total pressure is
\begin{equation*}
p =p _{B}+p _{M},
\end{equation*}
where $p_B$ is the baryonic pressure given by
\begin{equation}
p _{B}=\frac{\gamma }{2\pi ^{2}}\sum\limits_{j}\int_{0}^{k_{Fj} }\frac{p^{4}}{\sqrt{p^{2}+m _{j}^{\ast2 }}}dp,
\end{equation}

and $p_M$ is the pressure of the mesons:

\begin{eqnarray}
p _{M}&=&-\frac{\left( g_{\sigma N}\sigma _{0}\right) ^{2}}{%
2\Delta _{\sigma }}+\frac{\left( g_{\omega N}\omega
_{0}\right) ^{2}}{2\Delta _{\omega }}+\frac{\left( g_{\rho N}\rho
_{0}\right) ^{2}}{2\Delta _{\rho }} \nonumber \\
&&-\frac{\left( g_{\sigma N}\sigma^{\ast}_{0}\right)
^{2}}{2\Delta _{\sigma \sigma^{\ast}}}+\frac{\left( g_{\omega N}\phi
_{0}\right) ^{2}}{2\Delta _{\omega \phi }} \nonumber\\
&&-\frac{1}{3}bM_{n}\left( g_{\sigma N}\sigma _{0}\right) ^{3}-\frac{1}{4}c\left( g_{\sigma N}\sigma _{0}\right)^{4}.
\end{eqnarray}

\section{Lambdas in (Hyper)Nuclear Matter}\label{couplings}

\paragraph*{}
Inspired by the pioneer works on the role of the isospin in the liquid-gas phase transition \cite{ Muller95,Chomaz04,Ducoin06,Const00,Const01,Debora01,NPsp,Spinodal01}, 
along with more recent works on the role of the strangeness in the phase transition of dense neutron star matter \cite{Fran01,Fran02,Fran03,Fran04,Fran05,Fran06}, 
in this work we want to study the role of strangeness in the 
low density and zero temperature LG phase transition, which can be phenomenologically associated to multiple strange bound hypernuclei \cite{Mallik15,Khan15}.
\paragraph*{}
Because of the huge uncertainties in the strange sector we do not aim at having quantitative predictions on that phase transition, but would like to get qualitative
statements and avoid as much as possible the model dependence of the results.
For this reason we shall explore as widely as possible the largely unconstrained parameter space of the hyperon couplings. In this section we detail  the criteria
employed to fix the size of the parameter space.
\paragraph*{}  
Concerning the nucleon sector, we used the GM1 parameterization for the NLWM \cite{Moszkowski} and the original Walecka 
\cite{Walecka} parametrization for the LWM.
\paragraph*{}
The two sets of parameters are denoted by NLWM and LWM respectively shown in Tab.\ref{TabI} with the fitted nuclear bulk properties.

\begin{table}[ht]
\centering
\begin{tabular}{cccc}
\hline
&  {\bf NLWM}&{\bf LWM} \\
\hline
$n_0$ (fm$^{-3}$)             &   0.153    & 0.17\\
$K$ (MeV)                     &   300      & 554\\
$m^*/m$                       &   0.70     & 0.54\\
$B/A$ (MeV)                  &   -16.3     & -15.95\\
${\cal E}_{\rm sym}$ (MeV)    &   32.5     & 39.22\\
$L$ (MeV)                     &   94       & 127.22\\
\hline
$\Delta _{\sigma }$                    ($\text{fm}^{2}$)   &   11.785     & 13.670 \\
$\Delta _{\omega }$                    ($\text{fm}^{2}$)   &   7.148      & 10.250 \\
$\Delta _{\rho }$                      ($\text{fm}^{2}$)   &   4.410      & 4.410 \\
$\Delta _{\sigma {\sigma^{\ast} }}$  ($\text{fm}^{2}$)   &     3.216      & 3.769 \\
$\Delta _{\omega\phi }$                ($\text{fm}^{2}$)   &   4.212      & 6.040 \\
$b$                        & \;\; 0.002947  & \;\; 0.000 \;\; \\
$c$                        & -0.001070  & 0.000  \\
\hline
\end{tabular}
\caption{Sets of parameters used in this work and corresponding saturation properties.}
\label{TabI}
\end{table}

\paragraph*{}
It is well known that the value of the symmetric nuclear matter incompressibility  does not qualitatively influence the phase diagram, nor do the uncertainties on the other parameters.
We therefore consider the NLWM couplings as sufficiently well settled and do not play with them in the following. 
To fully explore the phenomenology of the model in the strange sector, the hyperon couplings are considered as free parameters, which however have to fulfill minimal requirements in 
terms of the potential and the hypernuclei data.
To be clear with the notation in the following, the general function associated 
to the $\Lambda$-potential is the three variable function: $\mathcal{U}_{\Lambda}\left(n_{n},n_{p},n_{\Lambda}\right)$. For symmetric matter $n_n=n_p$ we have a 
two variable function $U_{\Lambda}\left(n_{N},n_{\Lambda}\right)$.
The one variable $\Lambda N$ potential is denoted by $U_{\Lambda}^{N}\left(n_{N}\right)\equiv U_{\Lambda}\left(n_{N},n_{\Lambda}=0\right)$ and finally for $\Lambda \Lambda$ 
potential we have $U_{\Lambda}^{\Lambda}\left(n_{\Lambda}\right)\equiv U_{\Lambda}\left(n_{N}=0,n_{\Lambda}\right)$, where $n_N=n_n+n_p$ is the nucleon density.
For simplicity sometimes we omit the dependence of the potential function with respect to the density variables.
The $\chi_{\Lambda}$ couplings tell us how attractive or repulsive the $U_{\Lambda}$ can be. 
For the hyperon coupling constants, it is difficult to fix these phenomenological parameters due to the scarcity of data available in special for the multi-hyperon nuclei. 
Hence when the $\sigma^{\ast}$ and $\phi$ are taken into consideration we need some data from single-$\Lambda$ and double-$\Lambda$ nuclei.
Based on data on single-$\Lambda$ produced in $\left(\pi^{+},~K^{+}\right)$ reactions, the presently accepted value of the single-$\Lambda$ in symmetric nuclear matter 
at saturation density, $U_{\Lambda}^{N}\left(n_{0}\right)$
is $\approx-30$ MeV \cite{Hyper00_ULN,Hyper01_ULN}. For multi-hyperon there are available data just for the double-$\Lambda$ light nuclei, 
like $^{10}_{\Lambda\Lambda}$Be, $^{13}_{\Lambda\Lambda}$Be and $^{6}_{\Lambda\Lambda}$He, 
and the measurements are related to the $\Lambda\Lambda$ bond energy. This energy can be estimated from the binding energy difference between double-$\Lambda$ and 
single-$\Lambda$ hypernuclei denoted by $\Delta B_{\Lambda\Lambda}$.
In this work we consider the following value $\Delta
B_{\Lambda\Lambda}=0.67$ MeV
\cite{Aoki09,Ahn13,Hyper03_ULL,Hyper04_ULL,Hyper05_ULL}. The
$\Delta B_{\Lambda\Lambda}$ can be interpreted  as a rough estimation of the $U_{\Lambda}^{\Lambda}$
potential at the average $\Lambda$ density $\left\langle n_{\Lambda} \right\rangle \sim n_{0}/5$ inside the hypernucleus \cite{Hyper01_ULN}, where $n_{0}$ is the saturation 
point of symmetric nuclear matter in Table \ref{TabI}. 
Hence, the $U_{\Lambda}^{N}\left(n_{0}\right)=-28~\text{MeV}$ potential data can be used to tie the $\chi_{\omega \Lambda} $ to the $\chi_{\sigma \Lambda}$. 
For strange mesons, using $U_{\Lambda}^{\Lambda}\left(n_{0}/5\right)=-0.647~\text{MeV}$ we intend to link the $\chi_{\phi \Lambda} $ to the $\chi_{\sigma^{\ast} \Lambda}$. 
The general form of the $\Lambda$-potential $U_{\Lambda}$
in the RMF models considered is given by
\begin{eqnarray}
U_{\Lambda}\left(n_{N},n_{\Lambda}\right)&=&\chi_{\omega \Lambda}\left(g_{\omega N}\omega_{0}\right)+\chi _{\phi \Lambda}\left(g_{\omega N}\phi_{0}\right) \nonumber \\
&&-\chi_{\sigma \Lambda}\left(g_{\sigma N}\sigma_{0}\right)-\chi_{\sigma^{\ast} \Lambda}\left(g_{\sigma N}\sigma_{0}^{\ast}\right),
\label{UL_fileds}
\end{eqnarray}
where the dependence on the densities is given by the equations of motion of the meson fields, and $n_{N}$ is symmetric nuclear matter density. 
Nucleons and  $\Lambda$'s exchange $\sigma$ and $\omega$ mesons with each other, the first one being attractive while the second acts repulsively. These two mesons have no strange quantum number.
The additional strange mesons are similar to the ordinary $\sigma$ and $\omega$ but they see just the strange baryons, namely hyperons. 
The attractive force is due to the scalar meson $\sigma^{\ast}$ and the repulsive is due to the strange vector meson $\phi$. 
For simplicity we can define $\omega=\left(g_{\omega N}\omega_{0}\right)$, $\phi=\left(g_{\omega N}\phi_{0}\right)$, $\sigma=\left(g_{\sigma N}\sigma_{0}\right)$ and $\sigma^{\ast}=\left(g_{\sigma N}\sigma_{0}^{\ast}\right)$
to rewrite Eq.(\ref{UL_fileds}) in terms of the densities instead of the fields
\begin{eqnarray}
U_{\Lambda}\left(n_{N},n_{\Lambda}\right)&=&\chi_{\omega \Lambda}\left(\frac{g_{\omega n}}{m_{\omega}}\right)^2 n_{N}-\chi_{\sigma \Lambda}\left(\frac{g_{\sigma n}}{m_{\sigma}}\right)^2\rho^{s}_{N}\left(\sigma\right) \nonumber \\
&&+\left[1+\left(\frac{\chi_{\phi \Lambda}}{\chi_{\omega \Lambda}}\right)^2\left(\frac{m_{\omega}}{m_{\phi}}\right)^2\right]\nonumber \\
&&\times \left(\frac{g_{\omega n}}{m_{\omega}}\right)^2\left(\chi_{\omega \Lambda}\right)^2n_{\Lambda} \nonumber\\
&&-\left[1+\left(\frac{\chi_{\sigma^{\ast} \Lambda}}{\chi_{\sigma \Lambda}}\right)^2\left(\frac{m_{\sigma}}{m_{\sigma^{\ast}}}\right)^2\right] \nonumber \\
&&\times\left(\frac{g_{\sigma n}}{m_{\sigma}}\right)^2\left(\chi_{\sigma \Lambda}\right)^2\rho_{\Lambda}^{s}\left(\sigma,\sigma^{\ast}\right)\nonumber\\
&&-\chi_{\sigma \Lambda}\left(\frac{g_{\sigma n}}{m_{\sigma}}\right)^2\left[-bm_{n}\sigma^2-c\sigma^3\right].
\end{eqnarray}

\paragraph*{}
We can look for one dimensional potential $U_{\Lambda}^{N}\left(n_{N}\right)$, which is just the $\Lambda$- potential for a $\Lambda$ in nuclear symmetric matter and this potential is
a single variable function in nucleon density. It reads
\begin{eqnarray}
U_{\Lambda}^{N}\left(n_{N}\right)&=&\chi_{\omega \Lambda}\left(\frac{g_{\omega N}}{m_{\omega}}\right)^2 n_{N}-\chi_{\sigma \Lambda}\left(\frac{g_{\sigma N}}{m_{\sigma}}\right)^2\nonumber \\
&&\times\left[\rho^{s}_{N}\left(\sigma\right)-bm_{n}\sigma^2-c\sigma^3\right].
\label{Uln_contraint_densities}
\end{eqnarray}

\paragraph*{}
Here we can use the data for $U_{\Lambda}^{N}\left(n_{0}\right)=-28~\text{MeV}$. Solving the above expression to $\chi_{\omega \Lambda}$ and using the equation of motion of the fields we have
\begin{eqnarray}
\chi_{\omega \Lambda}=\frac{\chi_{\sigma \Lambda}\left.\sigma\right|_{N=n_{0}}-28~\text{MeV}}{\left.\omega\right|_{N=n_{0}}}.
\label{Uln_contraint_fields}
\end{eqnarray}

\begin{figure}[ht!]
\centering
{\includegraphics[width=0.35\textwidth]{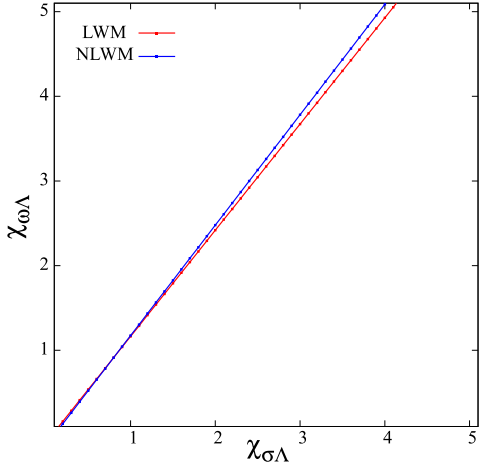}}
\caption{(Color online). Relations between parameters in RMF.}
\label{Figure 01}
\end{figure}

\paragraph*{}
The $\chi_{\sigma \Lambda}$ is left to be a free parameter in the RMF models. Fig.\ref{Figure 01} shows the relation between $\chi_{\sigma \Lambda}$ and $\chi_{\omega \Lambda}$ when we consider 
Eq.(\ref{Uln_contraint_fields}). For each choice of the
$\chi_{\sigma \Lambda}$, a particular potential is obtained in such a
way that it is constrained to $U_{\Lambda}^{N}\left(n_{0};\chi_{\sigma \Lambda}\right)=-28~\text{MeV}$.
The linear dependence obtained means that in the framework of the (N)LWM a 
strong attraction at low densities is always correlated to a strong repulsion at high densities. It is interesting to remark that the same is true in non-relativistic 
models \cite{Fran01,Fran02,Fran03,Fran04,Fran05}.

\begin{figure}[ht!]
\centering
\subfloat[]
{\includegraphics[width=0.43\textwidth]{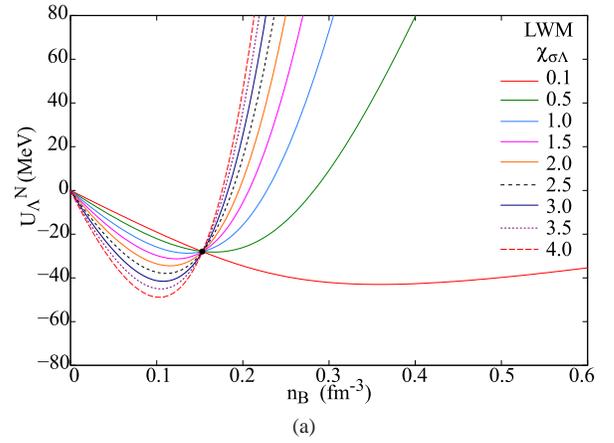}}

\subfloat[]
{\includegraphics[width=0.43\textwidth]{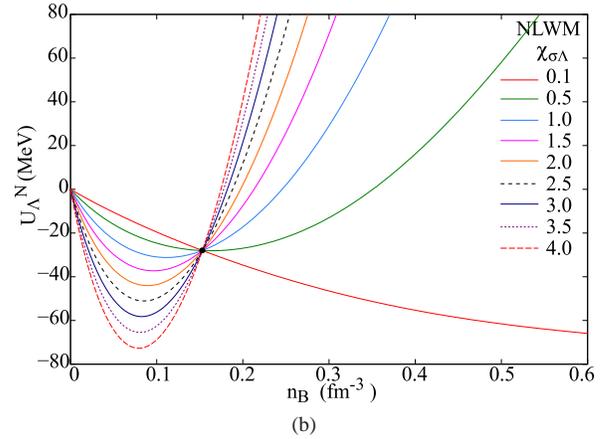}}

\caption{(Color online). $U_{\Lambda}^{N}$ curves (constrained by data $U_{\Lambda}^{N}\left(n_{0}\right)=-28$ MeV denoted by the black point) for some values of $\chi_{\sigma \Lambda}$ 
in (a) LWM  and (b) NLWM.}
\label{Figure 02}
\end{figure}

\paragraph*{}
Fig.\ref{Figure 02} shows the family of potentials constrained by Eq.(\ref{Uln_contraint_fields}) in LWM and NLWM. 
We can see that a very wide variety of behaviors is compatible with the hypernuclei
constraint, which explains why dedicated RMF works to hypernuclear structure 
have been able in the literature to reasonably 
fit the available single-particle levels with a large variety of choices for the couplings.
We can also observe that the LWM and NLWM models produce very similar behaviors for this potential.
The main difference between the two models, for large $\chi_{\sigma \Lambda}$, 
is that the $U_{\Lambda}^{N}$ potential in NLWM is deeper at low densities than the LWM  due to the nonlinear terms and the parameterization chosen.
\paragraph*{}
Now, we turn our attention to the $U_{\Lambda}^{\Lambda}\left(n_{\Lambda}\right)$ potential:

\begin{figure}[ht!]
\centering
\subfloat[]
{\includegraphics[width=0.43\textwidth]{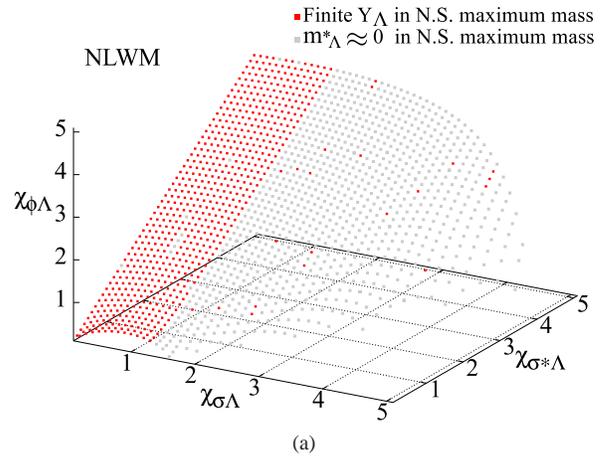}}

\caption{(Color online). Relations between parameters in the NLWM. 3D parameter space for for $\chi_{\phi \Lambda}$ constrained by $U_{\Lambda}^{\Lambda}$ potential. Gray points refer to parameters with which there is no numerical convergence in hyperonic stellar matter. }
\label{Figure 03}
\end{figure}

\begin{eqnarray}
U_{\Lambda }^{\Lambda }\left( n_{\Lambda }\right)&=&\left[ 1+\left(\frac{\chi_{\phi \Lambda}}{\chi_{\omega \Lambda}}\right)^{2}\left( \frac{m_{\omega }}{m_{\phi }}\right) ^{2}\right]\left( \chi_{\omega \Lambda}\right)\omega \nonumber \\
&&-\alpha\left(\chi_{\sigma \Lambda}\right)\Sigma,
\label{Ull_alpha}
\end{eqnarray}
where we have defined: $\alpha=1+\left(\frac{\chi_{\sigma^{\ast} \Lambda}}{\chi_{\sigma \Lambda}}\right) ^{2}\left( \frac{m_{\sigma }}{m_{\sigma ^{\ast }}}\right)^{2}$ and 
\begin{equation}
\Sigma=  
\sigma -\left(\frac{g_{\sigma N}}{m_{\sigma }}\right) ^{2}\left( \frac{\alpha-1}{\alpha }\right) \left(-bm_{n}\sigma ^{2}-c\sigma ^{3}\right).
\end{equation}%
The other chosen data is $U_{\Lambda}^{\Lambda}\left(\frac{n_{0}}{5}\right)=-0.67~\text{MeV}$. Therefore, solving Eq.(\ref{Ull_alpha}) for $\chi_{\phi \Lambda}$, we obtain:
\begin{eqnarray}
\chi_{\phi \Lambda}&=&\left( \frac{m_{\phi }}{m_{\omega }}\right) \nonumber\\
&&\times\sqrt{\frac{U_{\Lambda}^{\Lambda}\left(\frac{n_{0}}{5}\right)+\alpha \chi_{\sigma \Lambda}\left. \Sigma \right\vert _{n_{\Lambda }=\frac{%
n_{0}}{5}}-\chi_{\omega \Lambda}\left. \omega \right\vert _{n_{\Lambda }=%
\frac{n_{0}}{5}}}{\chi_{\omega \Lambda} \left. \omega \right\vert
_{n_{\Lambda }=\frac{n_{0}}{5}}}}\chi_{\omega \Lambda}.\nonumber \\
&&
\label{Ull_contraint_fields}
\end{eqnarray}%

\paragraph*{}
The above expressions are valid for the NLWM, and the LWM expression is obtained for $b=c=0$, when the  $\Sigma$ is reduced to the $\sigma$ field.
Fig.\ref{Figure 03} shows the 3D parameter space $\chi_{\sigma \Lambda}$ $\times$ $\chi_{\sigma^{\ast} \Lambda}$ $\times$ $\chi_{\phi \Lambda}$ where we consider 
Eq.(\ref{Ull_contraint_fields}). Note that in Eq.(\ref{Ull_contraint_fields}) there are
combinations of $\chi_{\sigma \Lambda}$ and $\chi_{\sigma^{\ast}\Lambda}$ that do not result in real solutions. 
This gives a first trivial limitation for the parameter values.
The residual parameter space, shown in Fig.\ref{Figure 03} for the NLWM, is still extremely large. Minimal constraints can be 
added requiring that convergent solutions are obtained in  hyperonic stellar matter (with all the baryon octet, electrons and muons
included) in $\beta$ equilibrium. $\chi_{\rho}=1.5$ is fixed, so that we guarantee that  $\Lambda$'s are the first hyperons to appear
and the (unconstrained) couplings to $\Sigma$ do not play a major role.
The gray points in Fig.\ref{Figure 03} are related to divergent solutions, where
the $\Lambda$ effective mass goes to zero at some finite density.  
The red points yield possible solutions and, in some cases, 
the maximum masses can reach two solar masses with a finite
$Y_{\Lambda}$ \cite{Luiz,Gusakov}.

\onecolumngrid

\begin{figure}[ht!]
\centering
\subfloat[]
{\includegraphics[width=0.45\textwidth]{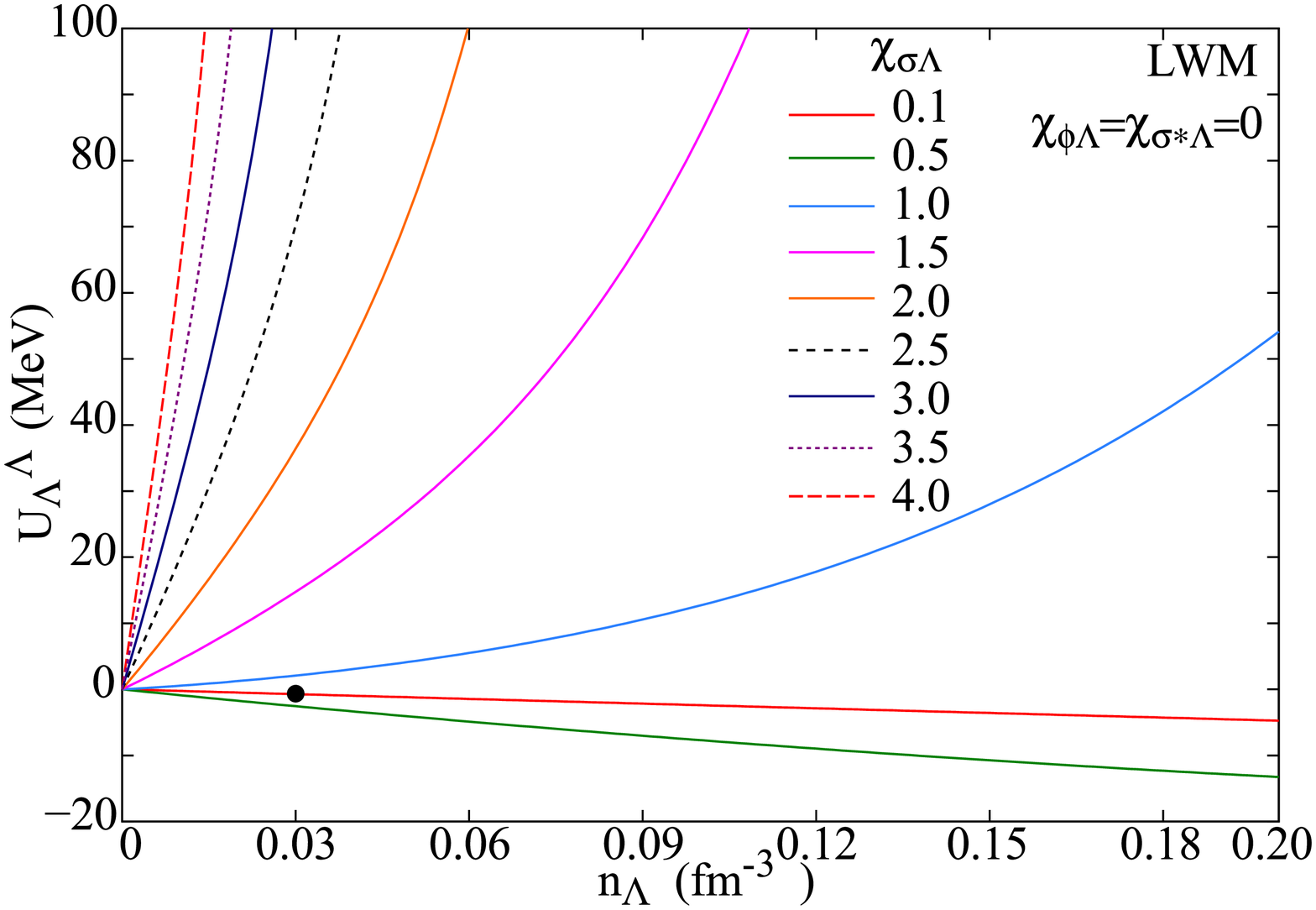}}
\subfloat[]
{\includegraphics[width=0.45\textwidth]{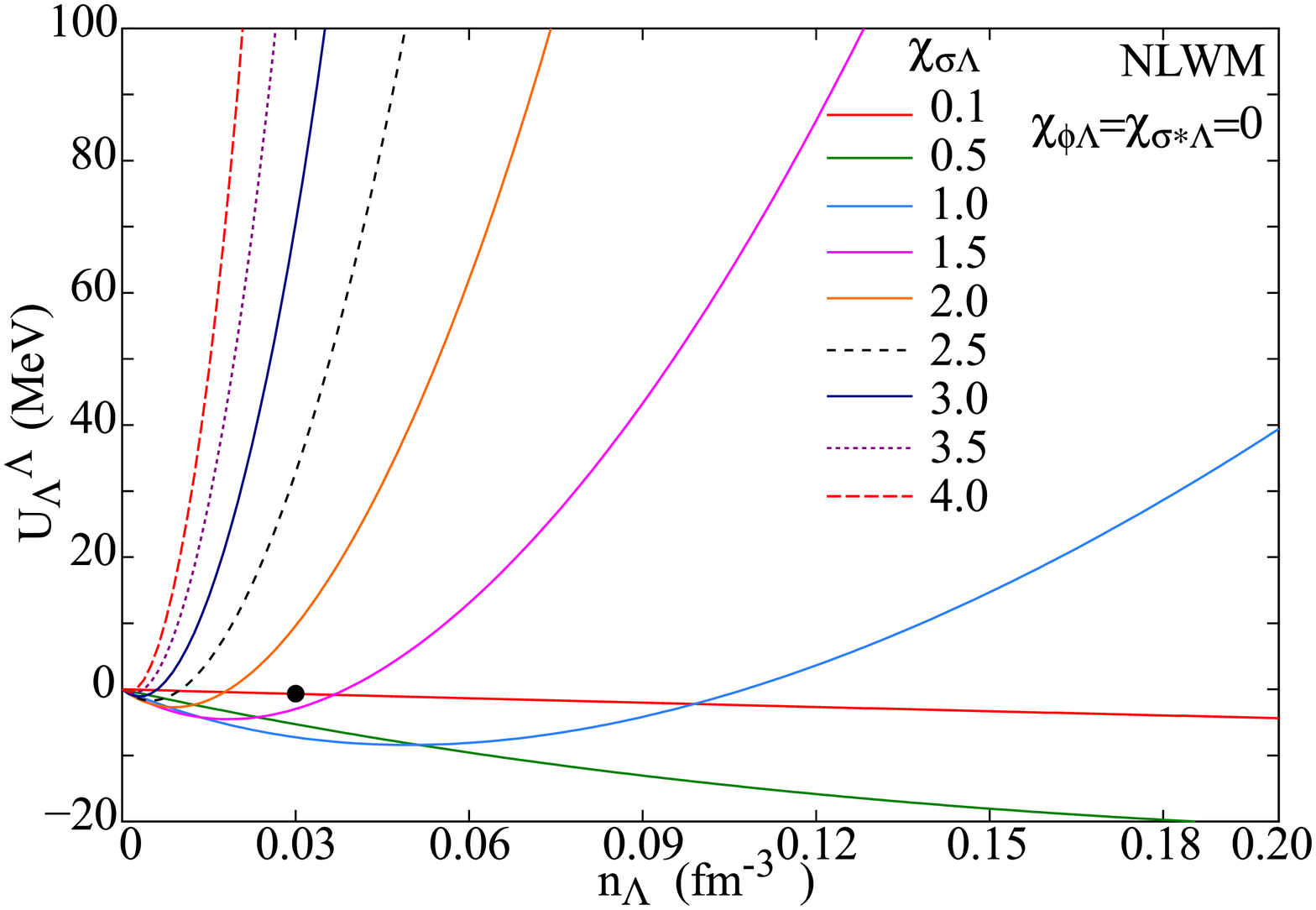}}

\subfloat[]
{\includegraphics[width=0.45\textwidth]{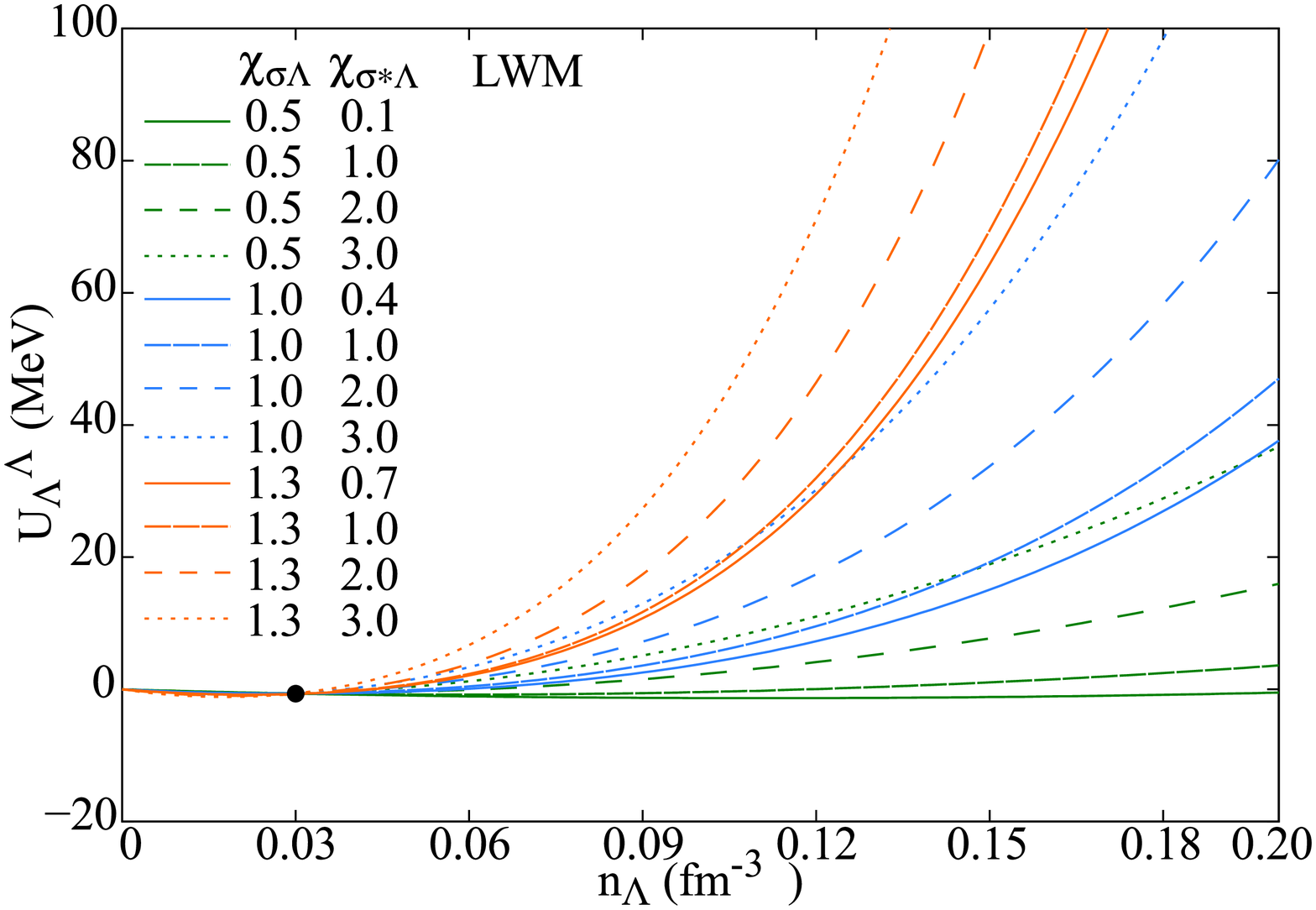}}
\subfloat[]
{\includegraphics[width=0.45\textwidth]{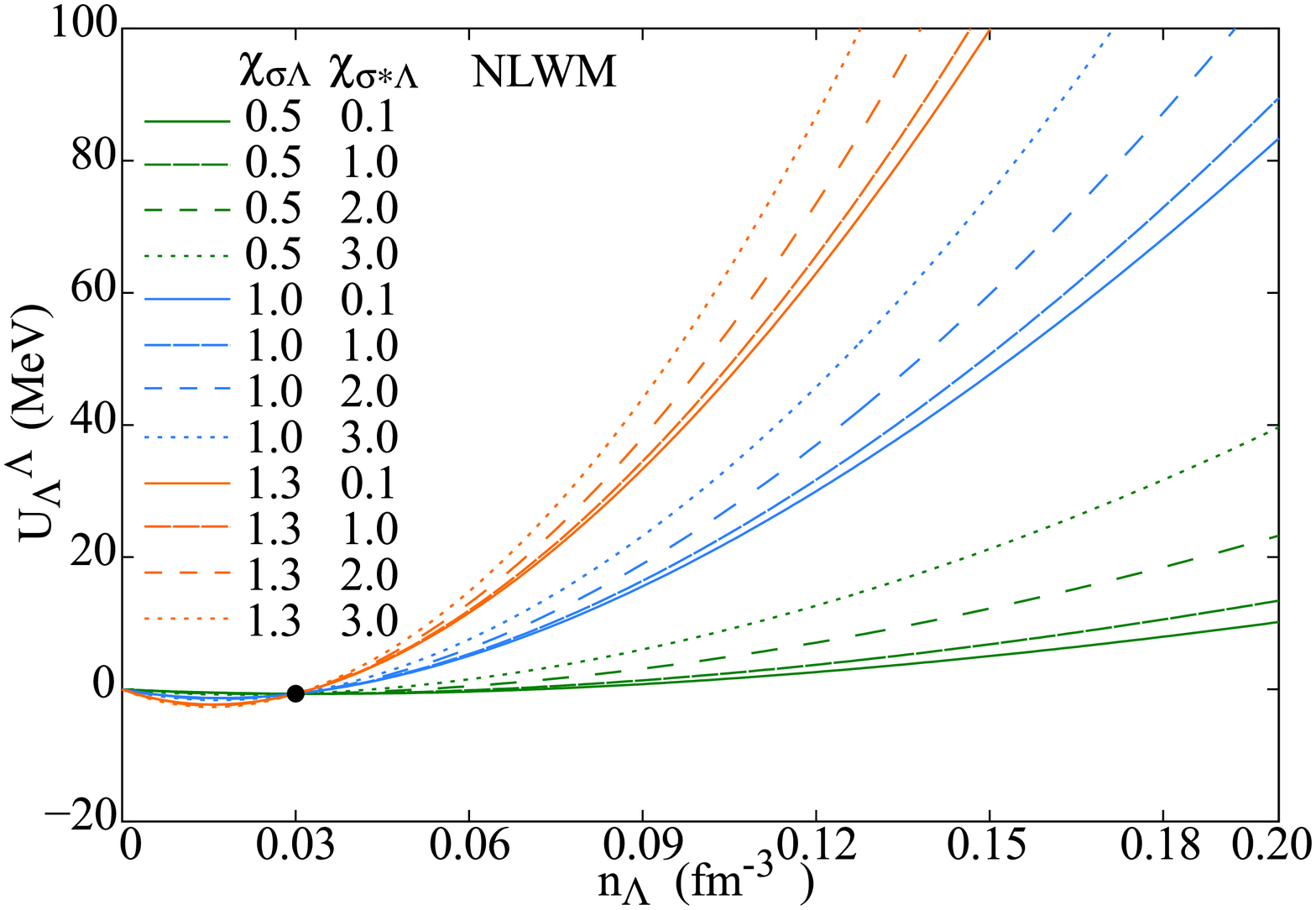}}

\caption{(Color online). The black points in each of these figure denote $U_{\Lambda}^{\Lambda}\left(n_{0}/5\right)=-0.67$ MeV. 
(a) and (b) show $U_{\Lambda}^{\Lambda}$ potential without strange mesons for some values of $\chi_{\sigma \Lambda}$ in LWM  and NLWM respectively. 
(c) and (d) show $U_{\Lambda}^{\Lambda}$ potential constrained to pass through the black point for some pairs of values of $\chi_{\sigma \Lambda}$ and $\chi_{\sigma^\ast \Lambda}$ in LWM  and NLWM respectively.}
\label{Figure 04}
\end{figure}

\twocolumngrid

This study is only done with the NLWM because it is well
known that the LWM leads to irrealistic results for high density matter.
Of course the LWM and NLWM gives different EOS even at low density, but for now it is enough to restrict our parameter
space substantially to start the study of possible instabilities in hypernuclear matter at low density. Later,
we see how drastic our choice for the $\chi_{\sigma \Lambda}$
parameter is when we discuss the instabilities. A very similar
reasoning without strange mesons was proposed in Ref.\cite{Glenn00}, where
experimental values of the  $U_{\Lambda}^{N}$ were used to restrict
the hyperon-meson coupling constants. In that paper, the resulting maximum stellar masses were
also analysed. 
Adding this condition still leaves us with a wide two-dimensional parameter space,
which corresponds to an almost unconstrained model. A major simplification 
would be obtained if we do not introduce extra strange mesons.
Indeed if we put $\chi_{\phi \Lambda}=\chi_{\sigma^{\ast}\Lambda}=0$
we are left with two equations and two unknowns, leading to a unique parameter choice for each of the models. This choice might sound appealing, especially if we
recall that historically strange mesons were added \cite{Schaffner94} to provide extra binding in the $\Lambda$-$\Lambda$ channel based on an analysis 
of hypernuclear data which nowadays appears questionable \cite{Shen06}. 

\paragraph*{}
The families of $U_{\Lambda}^{\Lambda}$ potential curves without strange mesons obtained with the LWM and the NLWM are shown in 
Figs.\ref{Figure 04} (a) and (b). We can see that the only possibility of having the very small extra binding suggested by experimental data, 
at the low densities explored in hypernuclei, is to have a potential which is unrealistically attractive at higher densities. This is due to the 
linear correlation between $\chi_{\sigma \Lambda}$ and $\chi_{\omega \Lambda}$ 
observed in Fig.\ref{Figure 01}. Consequently, the resulting EOS of stellar matter is clearly too soft. 
One can object that summarizing hypernuclear data to two values for the $\Lambda$ potential in infinite matter is a very crude approximation, which is certainly true. 
However it is well known from very different approaches that dedicated fits of hypernuclear data require some extra repulsion at higher density \cite{Millener88,Lonardoni14}, 
in qualitative agreement with our oversimplified nuclear matter reasoning. This discussion implies that a realistic RMF model should probably include strange mesons, 
or alternatively more complex non-linear couplings, even if this is done at the price of considerably enlarging the parameter space.
In particular in this paper, our motivation being to extract a phase diagram as general as possible, we prefer considering a parameter space which is too large to one 
which is too narrow. We will therefore stick to the parameter space defined by Fig.\ref{Figure 03}.

\paragraph*{}
Figs.\ref{Figure 04} (c) and (d) display the LWM and  NLWM $U_{\Lambda}^{\Lambda}$ potential with the inclusion of 
strange mesons, and with the extra requirement of fulfilling Eq.(\ref{Ull_contraint_fields}). We can see that a wide range of behaviors is still 
possible. Figs.\ref{Figure 04} (c) and (d) are globally 
similar, although in (d)  the potential is slightly deeper than in (c)
at very low density, i.e. $n_{\Lambda}<n_{0}/5$. For high densities,
we can clearly see that all curves in Fig.\ref{Figure 04} (d) are
steeper than in (c), ie, for the parameters chosen, the
$U_{\Lambda}^{\Lambda}$ is more attractive with the NLWM than with the
LWM. If one observes the values of the coupling constants, it is
obvious that as the  $\chi_{\sigma \Lambda}$ and $\chi_{\sigma^{\ast} \Lambda}$ 
values related to the attractive interactions increase, so do the 
$\chi_{\omega \Lambda}$ e $\chi_{\phi \Lambda}$ values, related to the
repulsive interaction.

\section{Results from AFDMC} \label{abinitio}
\paragraph*{}
In the recent years, ab-initio models based on the Brueckner or Dirac-Brueckner theory
\cite{Burgio11,Schulze11,Katayama14} or on different quantum Monte Carlo simulation techniques \cite{abinitio00,Carlson03,Gandolfi09,Gezerlis10,Gandolfi12,Gandolfi14} 
have been applied to (hyper)-nuclear matter. Such models provide in the pure neutron sector, in the low density regime where the underlying interactions are well known from scattering data and 
three-body effects are not expected to be important, a very essential constraint to phenomenological mean field models, which starts to be 
routinely applied in order to fix some of the unknown couplings.
Calculations including hyperons are still very scarce \cite{abinitio00,Burgio11,Schulze11,Katayama14}. We here compare our results to the very recent AFDMC model \cite{abinitio00},
which has been satisfactorily compared to hypernuclear data \cite{Lonardoni14} and  allows producing very massive neutron stars in agreement with the observations  \cite{abinitio00}, 
though with negligible strangeness fraction.
This model is based on a phenomenological bare interaction inspired
by the Argonne-Urbana forces \cite{Usmani08}, with the addition of a purely phenomenological three-body term. One of the advantages of the model is that the 
authors provide simple parametrizations of their numerical results for the neutron-$\Lambda$ energy functionals, allowing both an easier comparison with our RMF
results, and a straigthforward calculation of the instability properties of hyper-matter as predicted by an ab-initio model. This latter point  is discussed 
in the next section. The fit of the energy density of the neutron-$\Lambda$ mixture is given by \cite{abinitio00}:

\begin{eqnarray}
\epsilon _{\text{total}}\left( n_{n},n_{\Lambda }\right)  &=&\left[ a\left(\frac{n_{n}}{n_{0}}\right) ^{\alpha }+b\left( \frac{n_{n}}{n_{0}}\right)^{\beta }\right] n_{n} \nonumber \\
&&+\frac{1}{2m_{\Lambda }}\frac{3}{5}n_{\Lambda }\left(3\pi ^{2}n_{\Lambda }\right)^{2/3}\nonumber \\
&&+\left( m_{n}n_{n}+m_{\Lambda }n_{\Lambda }\right) \nonumber \\
&&+c_{1}^{^{\prime}}n_{\Lambda }n_{n}+c_{2}^{^{\prime }}n_{\Lambda }n_{n}^{2}.
\label{pederiva}
\end{eqnarray}%
In this expression, the first term represents the energy density of pure neutron matter, where the parameters $a$, $\alpha$, $b$ and $\beta$ 
are listed in Tab.\ref{TabII} and $n_{0}$ is saturation point of symmetric nuclear matter. 
The second term highlights the kinetic energy density of pure $\Lambda$-matter, and the last two terms,
obtained from the fitting of the Monte Carlo results for different $Y_{\Lambda}
=n_\Lambda/n_B$ fractions, provide an
analytical parametrization for the difference between Monte Carlo energies of pure $\Lambda$ and pure neutron matter.
Notice that $\Lambda$-$\Lambda$ interactions are neglected in Ref.\cite{abinitio00}, which explains why pure $\Lambda$ matter ($n_n=0$) 
behaves as a Fermi gas of noninteracting particles. This means that the extrapolations to high $\Lambda$ densities have to be considered with a critical eye.
The constants $c_{1}^{^{\prime }}\equiv c_{1}/n_{0}$ and $c_{2}^{^{\prime}}\equiv c_{2}/n_{0}^{2}$ with $c_{1}$ and $c_{2}$ are given in Tab.\ref{TabIII}. 
Using Eq.(\ref{pederiva}), the chemical potentials become:
\begin{eqnarray}
\mu _{n}\left( n _{n}\right) &=&a\left( \alpha +1\right) \left( \frac{n_{n}}{n_{0}}\right) ^{\alpha }+b\left( \beta +1\right) \left( \frac{n_{n}}{n_{0}}\right) ^{\beta }\nonumber \\
&&+m_{n}+c_{1}^{^{\prime }}n_{\Lambda }+2c_{2}^{^{\prime
}}n_{\Lambda }n_{n},
\end{eqnarray}%
and
\begin{equation}
\mu _{\Lambda }\left( n_{\Lambda }\right) =\frac{1}{2m_{\Lambda }}\left(3\pi^{2}n_{\Lambda }\right) ^{2/3}+m_{\Lambda}+c_{1}^{^{\prime }}n_{n}+c_{2}^{^{\prime }}n_{n}^{2}.
\end{equation}%
From thermodynamics we can also write the total pressure as follows:
\begin{eqnarray}
p_{\text{total}}\left( n_{n},n_{\Lambda }\right)  &=&\left\{ \alpha a\left( 
\frac{n_{n}}{n_{0}}\right) ^{\alpha }+\beta b\left( \frac{n_{n}}{n_{0}}%
\right) ^{\beta }\right\} n_{n} \nonumber \\
&&+\frac{1}{5m_{\Lambda }}n_{\Lambda }\left( \frac{6\pi ^{2}n_{\Lambda }}{%
2s_{\Lambda }+1}\right) ^{2/3} \nonumber \\
&&+c_{1}^{^{\prime }}n_{n}n_{\Lambda }+2c_{2}^{^{\prime }}n_{\Lambda}n_{n}^{2}.
\end{eqnarray}

\paragraph*{}
In Tab.\ref{TabIII} we show the sets of parameters proposed by the authors of Ref.\cite{abinitio00} when only two-body forces are taken into account 
($\Lambda N$), and also with the consideration of  three body forces
that yield
two different parameterizations $\Lambda NN$ (I) and $\Lambda NN$ (II). 
In the case of pure neutron matter, in the AFDMC approach the binding energy has no free parameters and we can compare this result with the binding energy coming from our phenomenological RMF models. 
When we include a $\Lambda$-fraction in the system, the ab-initio model itself 
needs phenomenological inputs and is associated to theoretical error bars.
This is due to the need of three-body forces in order to properly reproduce hypernuclear data \cite{Lonardoni14}.
The interval of predictions between $\Lambda NN$ (I) 
and $\Lambda NN$ (II), obtained using two different prescriptions for the three-body force, will be interpreted in the following as the present 
theoretical error bar on ab-initio models, such that a phenomenological model like our RMF should lay between these two extreme cases.
\onecolumngrid

\begin{figure}[ht!]
\centering
\subfloat[]
{\includegraphics[width=0.45\textwidth]{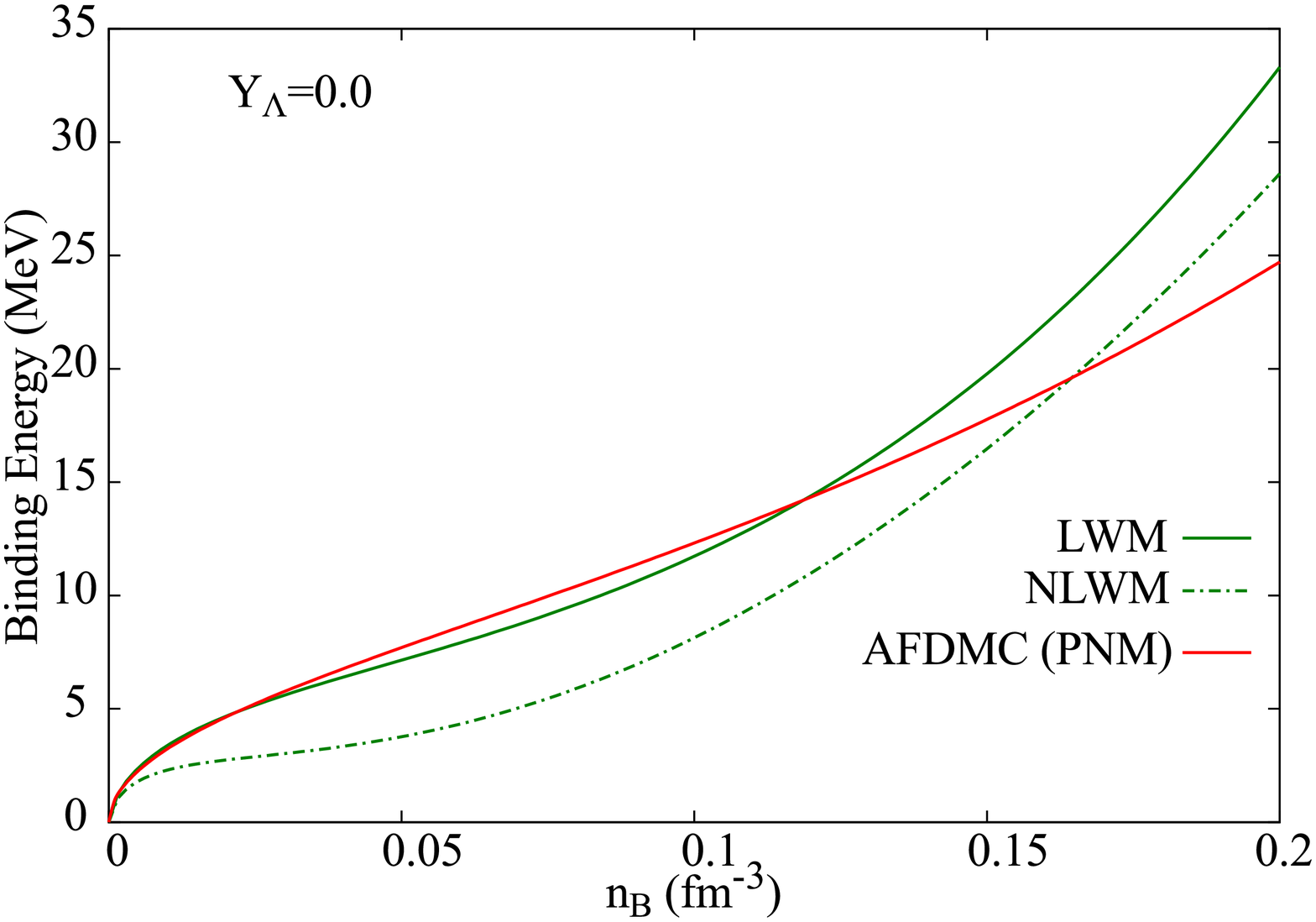}}
\subfloat[]
{\includegraphics[width=0.45\textwidth]{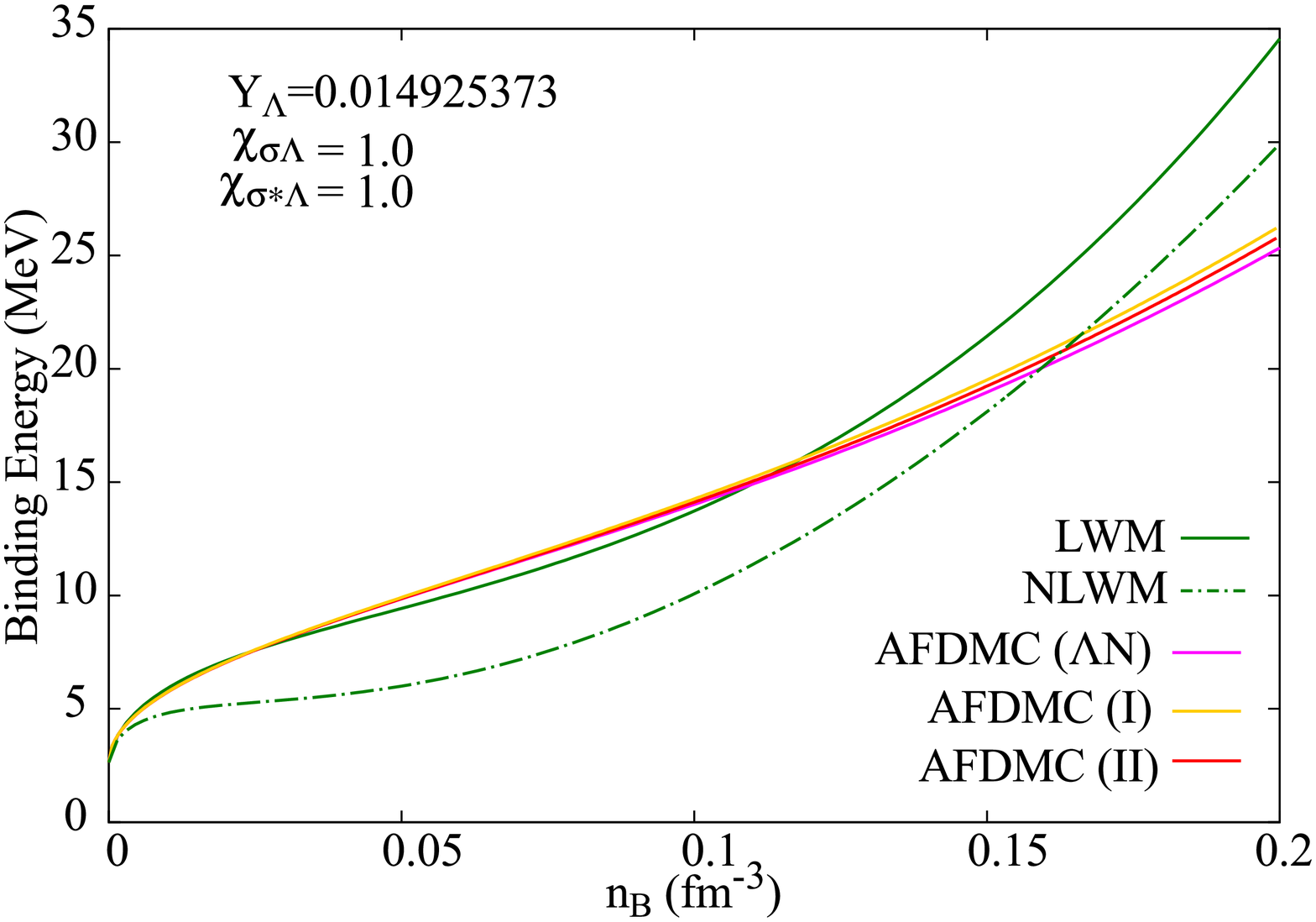}}

\subfloat[]
{\includegraphics[width=0.45\textwidth]{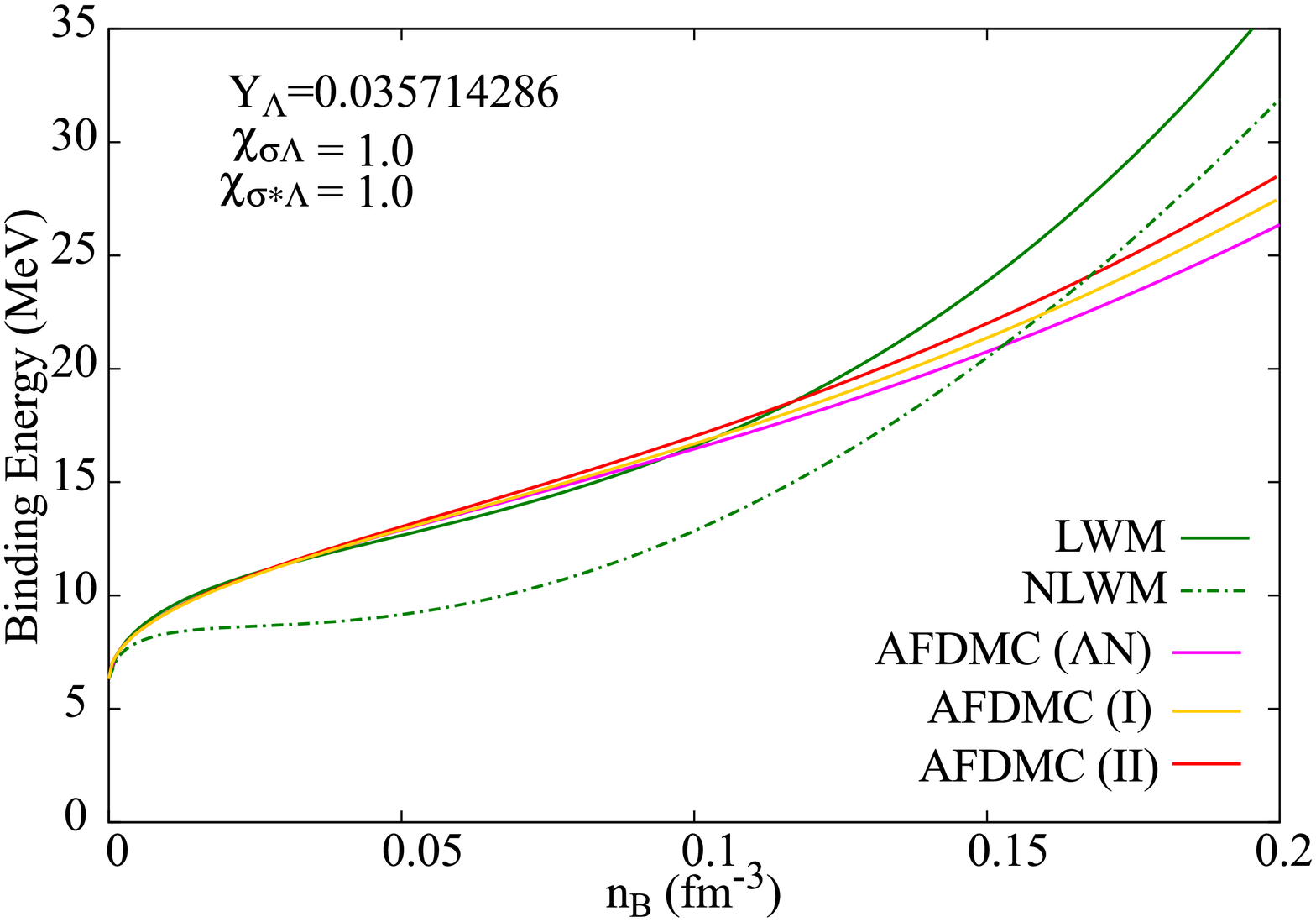}}
\subfloat[]
{\includegraphics[width=0.45\textwidth]{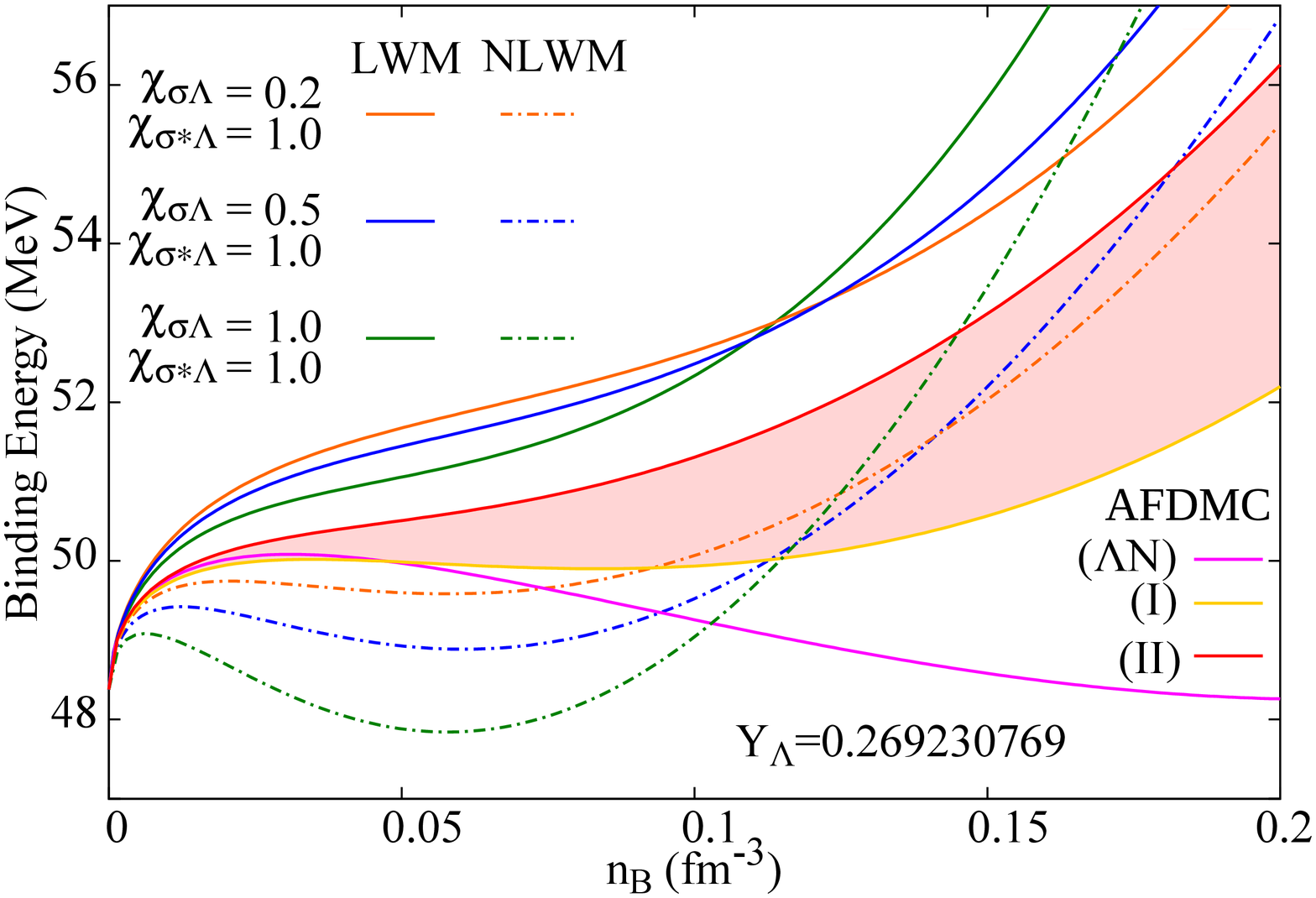}}

\caption{(Color online). Binding energy obtained with three different models AFMDC, LWM and NLWM, for different $\Lambda$-fractions show in (a), (b), (c) e (d) figures.}
\label{Figure 05}
\end{figure}

\twocolumngrid

\begin{table}[ht!]
\centering
\begin{tabular}{llll}
&&\\
PNM                                     \\
\hline
\hline
$n_0$ (fm$^{-3}$)             &   0.16  \\
$a$ (MeV)                     &   13.4  \\
$\alpha$                      &   0.514 \\
$b$ (MeV)                     &   5.62  \\
$\beta$                       &   2.436 
\end{tabular}
\caption{Set of parameters used in the AFDMC {ab-initio} model for PNM, from \cite{abinitio00}.}
\label{TabII}
\end{table}

\begin{table}[ht!]
\centering
\begin{tabular}{lll}
&&\\
$\Lambda N$                             \\
\hline
\hline
$c_{1}$ (MeV)                 &   -70.1 \\
$c_{2}$ (MeV)                 &     3.4 \\
&&\\
$\Lambda N+\Lambda NN (I)$               \\
\hline
\hline
$c_{1}$ (MeV)                 &   -77.0 \\
$c_{2}$ (MeV)                 &    31.3 \\
&&\\
$\Lambda N+\Lambda NN (II)$            \\
\hline
\hline
$c_{1}$ (MeV)                 &   -70.0 \\
$c_{2}$ (MeV)                 &    45.3 \\
\end{tabular}
\caption{Set of parameters used in the ab-initio AFDMC model including two and three body forces, from \cite{abinitio00}.}
\label{TabIII}
\end{table}

\paragraph*{}
In Fig.\ref{Figure 05} we plot the binding energy for different
values of the $\Lambda$-fraction present in Ref.\cite{abinitio00} for AFDMC and for representative RMF models. 
Fig.\ref{Figure 05}
(a) shows the binding energy for pure neutron matter.
It is known since a long time that RMF models are systematically too stiff at high
neutron density in comparison to ab-initio models. However
we can see that for the sub-saturation densities of interest for the present paper, 
the LWM agrees very well  with the
AFDMC, better than the NLWM, which in principle should be more sophisticated.  This remains true for finite $\Lambda$-fraction, as shown in Figs.\ref{Figure 05} (b) and (c), if this latter is small enough.
In this regime, the values of the $\Lambda$ coupling do not play an important role, and the same level of reproduction is obtained for different 
choices of  $\chi_{\sigma \Lambda},\chi_{\sigma^{\ast} \Lambda}$.
\paragraph*{}
The effect of three-body forces increases with increasing $\Lambda$-fraction,
and consequently the three versions of the AFDMC calculation start to
considerably  deviating from each other at the highest $\Lambda$-fraction
considered by the authors of \cite{abinitio00} (Fig.\ref{Figure 05} (d)).
In this condition, the AFDMC
($\Lambda N$) becomes very bound, 
due to the attractive feature of the $\Lambda N$ potential, while
the three-body force in 
AFMDC (I) and (II) insures the necessary repulsion to sustain 
massive neutron stars. We can see that at high $\Lambda$ fraction 
NLWM better reproduces the ab-initio results,  and the 
best reproduction is obtained for  $\chi_{\sigma \Lambda} \ll \chi_{\sigma^{\ast} \Lambda}$.
We have observed that $U_{\Lambda}^{\Lambda}$ is more sensitive to changes in $\chi_{\sigma \Lambda}$ than $\chi_{\sigma^{\ast} \Lambda}$ as seen in Figs.\ref{Figure 04} 
(c) and (d). No matter how much we change these parameters, we  
do not notably change the degree of agreement between the RMF models and the AFDMC. In this sense the orange and green curves in Fig.\ref{Figure 05} (d) represent
extreme choices for the RMF couplings in the two versions LWM (full lines) and NLWM (dashed lines).
To conclude, the inclusion of strange mesons is necessary to produce a RMF energy functional compatible with ab-initio results at low baryonic density.
For very low $\Lambda$-fractions, as it is the case in hypernuclei, the sensistivity 
to the $\Lambda$ couplings is very small, and the LWM surprisingly
leads to a very good agreement to the AFDMC parametrization.
However neither the linear nor the non-linear version of the WM are satisfactory, if one wants to describe matter with a non-negligible proportion of $\Lambda$'s,
and a dedicated fit with density dependent couplings should be done to
reduce the parameter space. For the purpose of the present paper we 
will continue with both models in our further analysis, keeping in mind that LWM results well reproduce ab-initio pure neutron matter, while NLWM with low
values of $\chi_{\sigma \Lambda}\approx 0.2-0.5$ should give a reasonably realistic description of symmetric and asymmetric matter with an important contribution of strangeness.

\section{Spinodal and Curvature Matrix} \label{spinodal}

\paragraph*{}
In the present section we focus on the calculation of the
instabilities in a system with neutrons, protons and  $\Lambda$'s at $T=0$ \cite{Fran01,Fran02,Fran03,Fran04,Fran05,Fran06}. 
A first order phase transition is signaled by an instability or concavity anomaly in the mean-field thermodynamic total energy density.
The total energy density of threee-component system is a three variable function of the densities. Therefore, we need to introduce the curvature matrix $\mathbf{C}$ associated to the scalar function $\epsilon$ 
at a point denoted by $P\in\left(n_{n}\times n_{p}\times n_{\Lambda}\right)$ .
Since our benchmark ab-initio model only contains neutrons and  $\Lambda$'s,
we consider first a two-component system case, where $P\in\left(n_{n}\times n_{\Lambda}\right)$, and later we comment about three-component systems \cite{Fran03,Fran06} 
which are more relevant for hypernuclear physics.
If $\epsilon$ is smooth, or at least twice continuously differentiable, $\mathbf{C}$ is symmetric. The curvature matrix elements are just second derivatives of the total energy density with 
respect to each independent variable. In our case the curvature matrix is just $2\times2$ matrix with elements: \cite{DG}

\begin{equation}
C_{ij}=\left( \frac{\partial ^{2}\epsilon \left( n_{i},n{j}\right) }{\partial n_{i}\partial n_{j}}\right)
\text{,}
\end{equation}%
where $i,j=n,\Lambda$. 
As this matrix is self-adjoint we can associate with it  one bilinear form and one quadratic form at point $P$.
So, the characteristic equation is

\begin{equation}
\mathtt{Det}\left( \mathbf{C-\lambda 1}_{2}\right) =0,
\end{equation}%
where $\mathbf{1}_{2}$ is $2\times2$ identity matrix. In another way,
\begin{equation}
\lambda ^{2}-\mathtt{Tr}\left( \mathbf{C}\right) \lambda +\mathtt{Det}\left(\mathbf{C}\right) =0.
\end{equation}%
The eigenvalues and eigenvectors of $\mathbf{C}$ have geometric meaning if $P$ is a critical point. We can solve their roots explicitly
\begin{equation}
\lambda _{1 }=\frac{1}{2}\left(\mathtt{Tr}\left( \mathbf{C}\right)+\sqrt{\mathtt{Tr}\left( \mathbf{C}\right) ^{2}-4\mathtt{Det}\left( \mathbf{C}\right) }\right)
\end{equation}%
and
\begin{equation}
\lambda _{2 }=\frac{1}{2}\left( \mathtt{Tr}\left( \mathbf{C}\right)-\sqrt{\mathtt{Tr}\left( \mathbf{C}\right) ^{2}-4\mathtt{Det}\left( \mathbf{C}\right) }\right),
\end{equation}%
where $\mathtt{Det}\left( \mathbf{C}\right)=\lambda _{1}\lambda _{2}$ and $\mathtt{Tr}\left( \mathbf{C}\right)=\lambda _{1}+\lambda _{2}$.  
The unitary eigenvectors are given by ${\widehat{n}^{1}}=\left(\delta n^{1}_{n},\delta n^{1}_{\Lambda}\right)$ and ${\widehat{n}^{2}}=\left(\delta n^{2}_{n},\delta n^{2}_{\Lambda}\right)$. 
For further analysis we define the direction by the ratios
\begin{equation}
\tan\theta_{1}=\frac{\delta n^{1}_{\Lambda}}{\delta n^{1}_{n}}=\frac{\lambda_{1}-C_{nn}}{C_{n\Lambda}}~~~\text{and}~~~\tan\theta_{2}=\frac{\delta n^{2}_{\Lambda}}{\delta n^{2}_{n}}=\frac{\lambda_{2}-C_{nn}}{C_{n\Lambda}},
\label{ratios_general}
\end{equation}%
where $\theta_{1}$ and $\theta_{2}$ are an angle measured counterclockwise from the positive $n_{n}$ axis. If $P$ is a critical point and hence $\mathbf{C}$ is just
a Hessian matrix so the determinant term is exactly the Gauss curvature and the trace is twice the mean curvature  \cite{DG}
\begin{equation}
K=\lambda _{1}\lambda _{2}~~\text{and}~~H=\frac{1}{2}\left(\lambda _{1}+\lambda _{2}\right).
\end{equation}%
The stability properties of the system depend on
the signs of the curvatures, $K$ and $H$, at each point $P\in\left(n_{n}\times n_{\Lambda}\right)$ \cite{Chomaz04,Ducoin06,Fran01,Fran02}:

\begin{enumerate}
 \item If $K>0$ and $H>0$, the system is stable. 
 \item If $K>0$ and $H<0$, the system is unstable, both eigenvalues are negative
and two independent order parameters should be considered meaning that more than two phases can coexist.
 \item If $K<0$, the system is unstable, meaning that the order parameter of the transition is always one-dimensional, 
similar to the nuclear liquid-gas phase transition at subsaturation densities.
 \item If $K=0$ and $H>0$, the system is stable. 
 \item If $K=0$ and $H<0$, the system is unstable. 
\end{enumerate}
In geometric terms the first and second condition tell us that $P$ represents an elliptic point, third a hyperbolic point and fourth and fifth a parabolic.
For a three-component system we have to calculate numerically the following equation
\begin{equation}
\mathtt{Det}\left( \mathbf{C-\lambda 1}_{3}\right) =0,
\end{equation}%
where $\mathbf{1}_{3}$ is a $3\times3$ identity matrix. In terms of the polynomials
\begin{equation}
\lambda ^{3}-\mathtt{Tr}\left( \mathbf{C}\right) \lambda ^{2}+\frac{1}{2}%
\left[ \mathtt{Tr}\left( \mathbf{C}\right) ^{2}-\mathtt{Tr}\left( \mathbf{C}%
^{2}\right) \right] \lambda -\mathtt{Det}\left( \mathbf{C}\right) =0,
\end{equation}%
where we have to analyse the signs of three eigenvalues. 
The remarkable feature of the liquid-gas phase transition is that one of all eigenvalues is negative and the associated eigenvector gives the instability direction \cite{Fran06}, 
what means that the energy surface is of a hyperbolic kind. Therefore, in the case of the simpler n$\Lambda$ system for a negative eigenvalue the ratio (\ref{ratios_general}) became
\begin{equation}
\frac{\delta n^{-}_{\Lambda}}{\delta n^{-}_{n}}=\frac{\lambda_{-}-C_{nn}}{C_{n\Lambda}}.
\label{ratios_negative nl}
\end{equation}%
The next equation will be useful in the discussion about the ratio for np$\Lambda$ system with symmetric condition $n_{n}=n_{p}$.
\begin{equation}
\frac{\delta n^{-}_{\Lambda}}{\delta n^{-}_{N}}=\frac{\lambda_{-}-C_{NN}}{C_{N\Lambda}}.
\label{ratios_negative npl}
\end{equation}%
\paragraph*{}
In the next section we comment our results.

\section{Results} \label{results}

\paragraph*{}
In order to understand the instabilities possibly present in the
models discussed in sections III and IV, we need 
the analysis done in the last section.
We have calculated the curvature matrix with the ab-initio and RMF models. The whole density space is a three-dimensional space and the spinodal region, when it occurs, 
is a three-dimensional volume that represents a geometric
\textit{locus} associated with the presence, at least, of one
negative eigenvalue. 
It is well known that in two-component systems with
neutrons and protons the liquid-gas phase transition occurs. The
corresponding two-dimensional spinodal zone appears below the
saturation density. So, in this system, one of the eingenvalues is negative.
For a more complex system, with neutrons, protons and  $\Lambda$'s for example, we can fix the $\Lambda$-fraction to see how the two-dimensional 
spinodal region in the neutron-proton plane changes when lambdas are added. 
In all the models analyzed, for any proton fraction, and with all the different choices of couplings, we have systematically found one and only one negative eigenvalue 
in a finite density space defining a spinodal region. The only exception is given by the n$\Lambda$ system studied with the LWM, which does not present any instability.
However the instability is there in the ab-initio model, and it appears in the LWM as soon as a non-zero proton fraction is added to the system, meaning that the result
of the LWM n$\Lambda$ mixture appears rather marginal. 

\onecolumngrid

\begin{figure}[ht!]
\centering
\subfloat[]
{\includegraphics[width=0.32\textwidth]{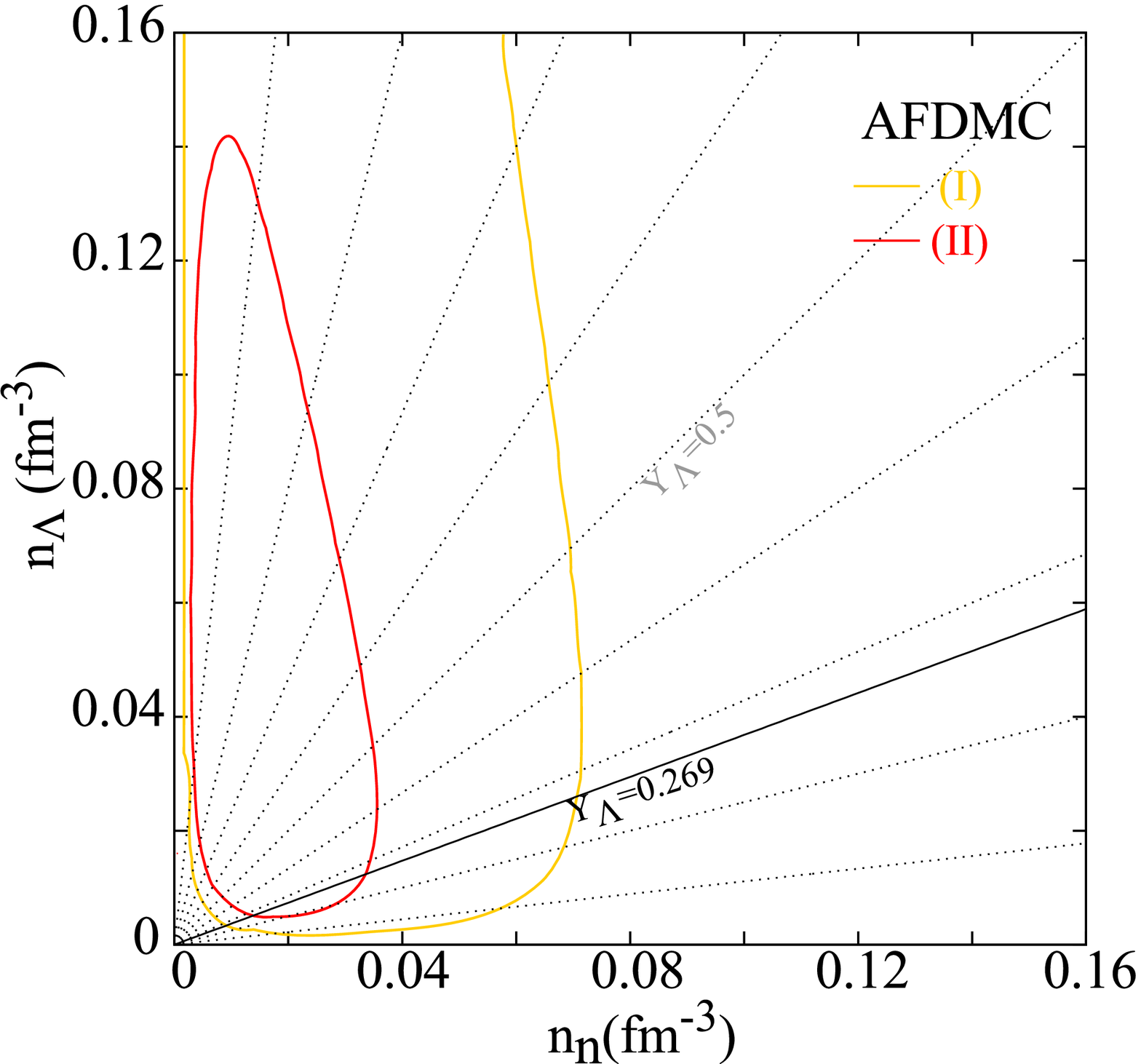}}
\subfloat[]
{\includegraphics[width=0.32\textwidth]{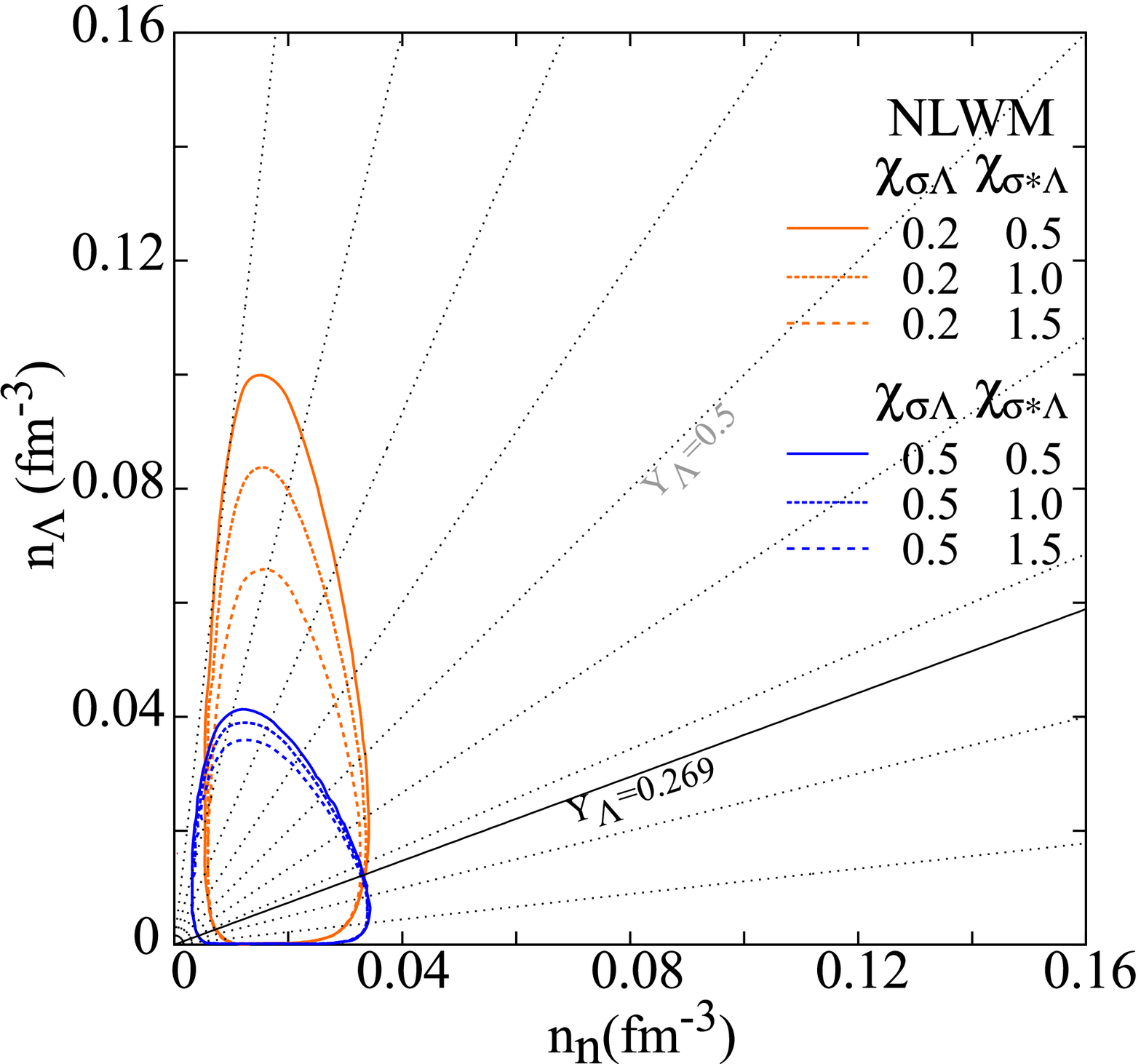}}
\subfloat[]
{\includegraphics[width=0.32\textwidth]{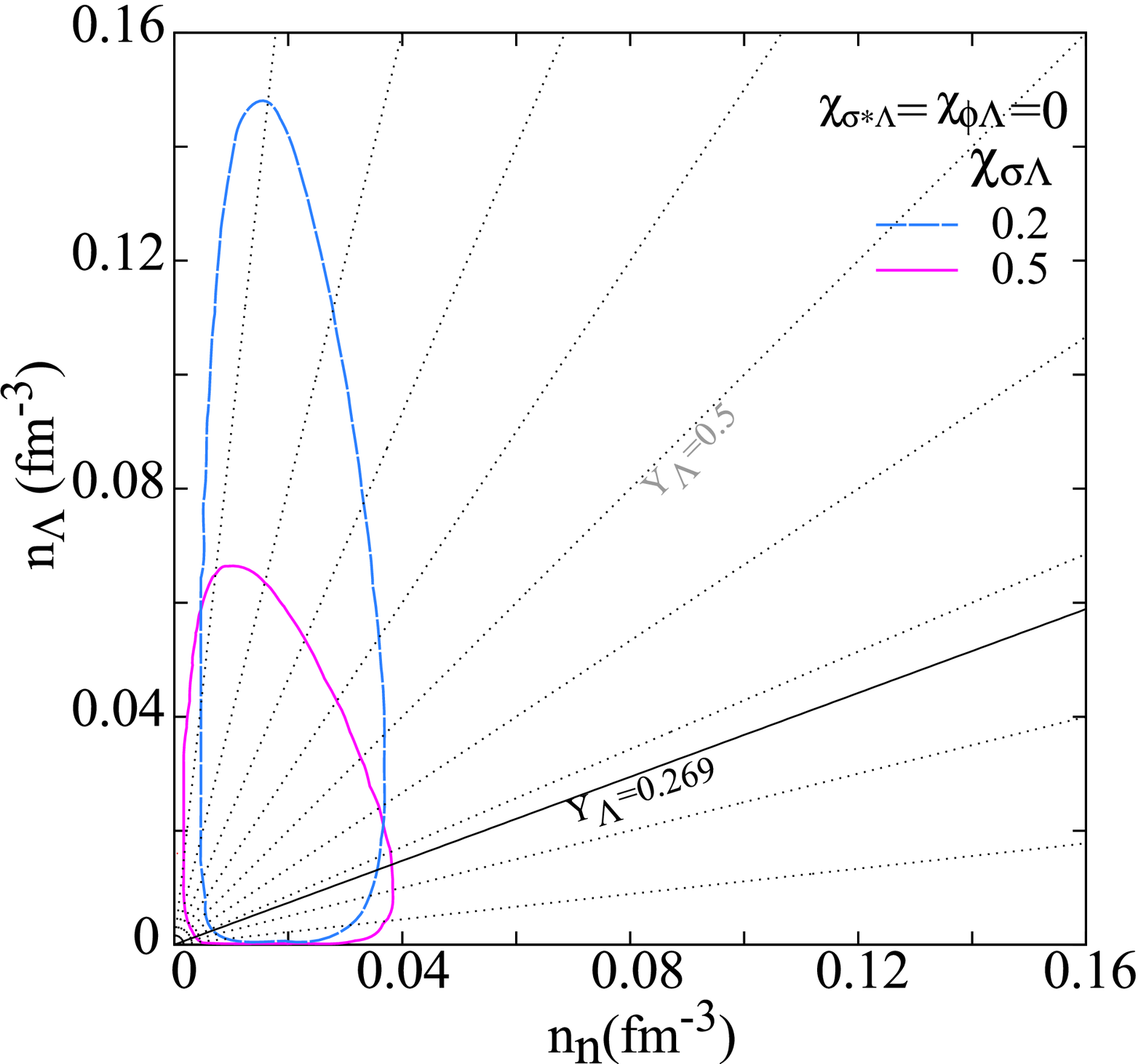}}

\caption{(Color online). Spinodals in the neutron-lambda plane for AFMDC with different parameterizations and NLWM with and without strange mesons.}
\label{Figure 06}
\end{figure}

\twocolumngrid

\paragraph*{}
Therefore we can conclude that a transition exists in the subsaturation nuclear matter including $\Lambda$ hyperons, and 
this transition belongs to the Liquid-Gas universality class.
In the following, we turn to study the characteristics of this transition in further details.

\paragraph*{}
In Fig.\ref{Figure 06} we plot the spinodal areas in a system
containing only neutrons and  $\Lambda$'s. In Fig.\ref{Figure 06} (a) two spinodal zones for the two
different parameterizations of the ab-initio model including three-body forces are shown. 
The behavior at high $\Lambda$-density should be considered with caution, since the AFDMC calculations were only done for $Y_\Lambda<0.269$.
In Fig.\ref{Figure 06} (b), different spinodal zones are shown for the NLWM taking into account different values of the strange mesons coupling constants. 
Two more spinodal curves for neutron-$\Lambda$ matter without strange mesons are also displayed in Fig.\ref{Figure 06} (c).
Note that none of these shapes touch the horizontal axis nor the
vertical one, even if they look very close to the $n_n$ axis for some of the models. This result is due to the fact that pure neutron and
pure $\Lambda$-matter are unbound. Indeed the spinodal instability 
at zero temperature leads to a phase transition where the system splits into two phases, the dense one representing the bound ground state. In the absence 
of a bound ground state, it is thus normal that the instability disappears.
In the following, whenever the spinodal zone
does not touch the axis it is clear that the reason underlying this
behavior is an unbound system.
It is interesting to observe that the widest extension of the instability is obtained with the most repulsive model. 
\onecolumngrid

\begin{figure}[ht!]
\centering
\subfloat[]
{\includegraphics[width=0.32\textwidth]{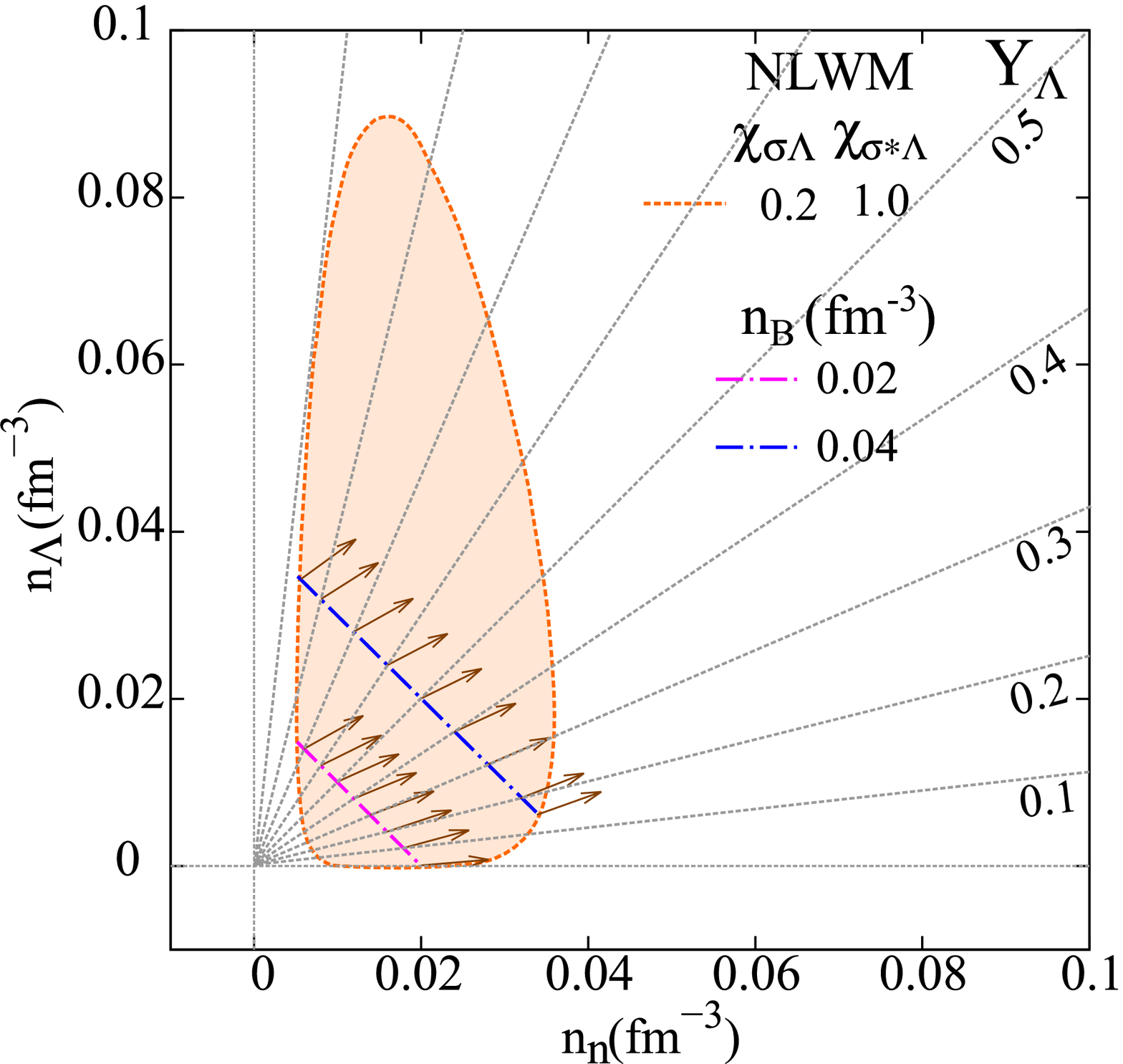}}
\subfloat[]
{\includegraphics[width=0.32\textwidth]{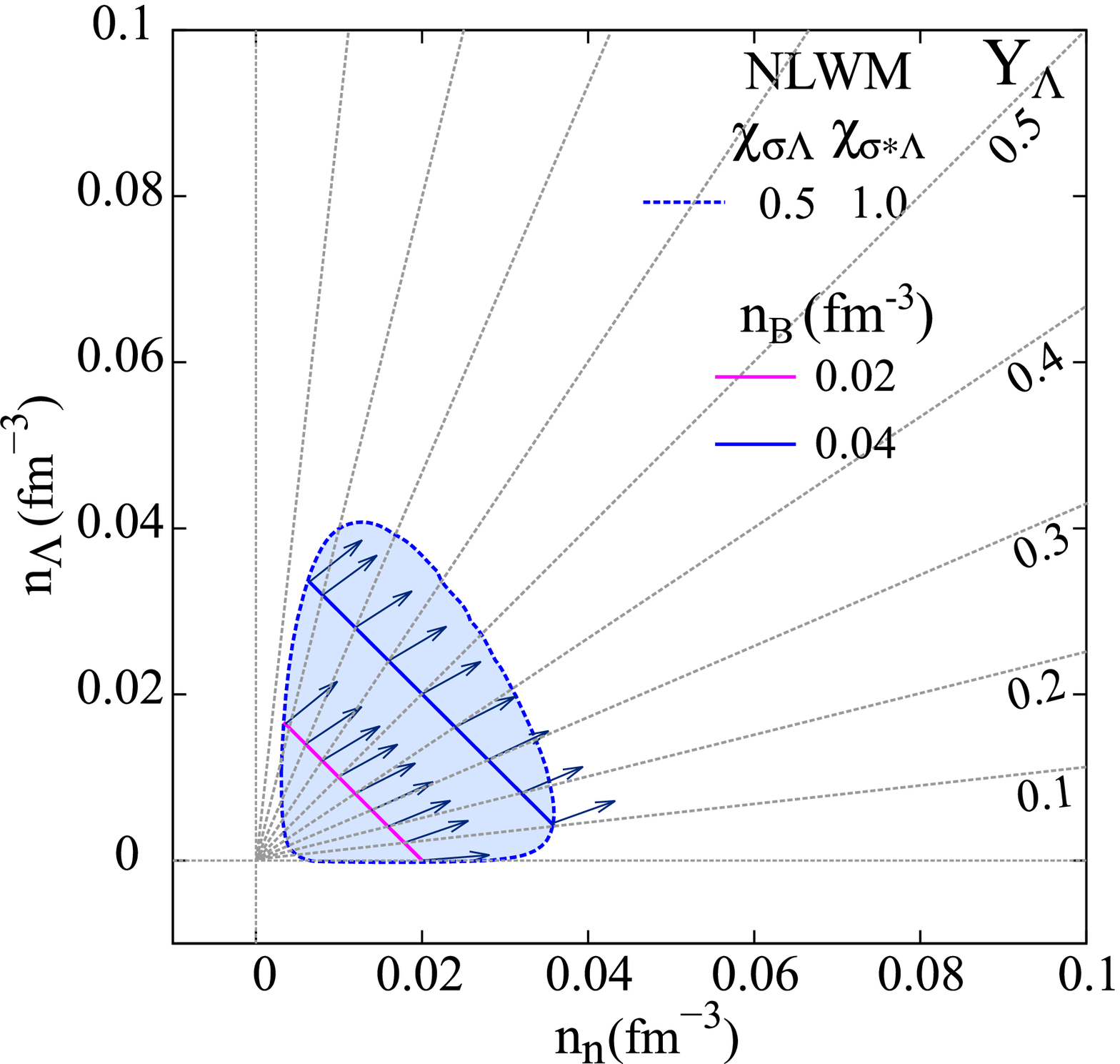}}
\subfloat[]
{\includegraphics[width=0.32\textwidth]{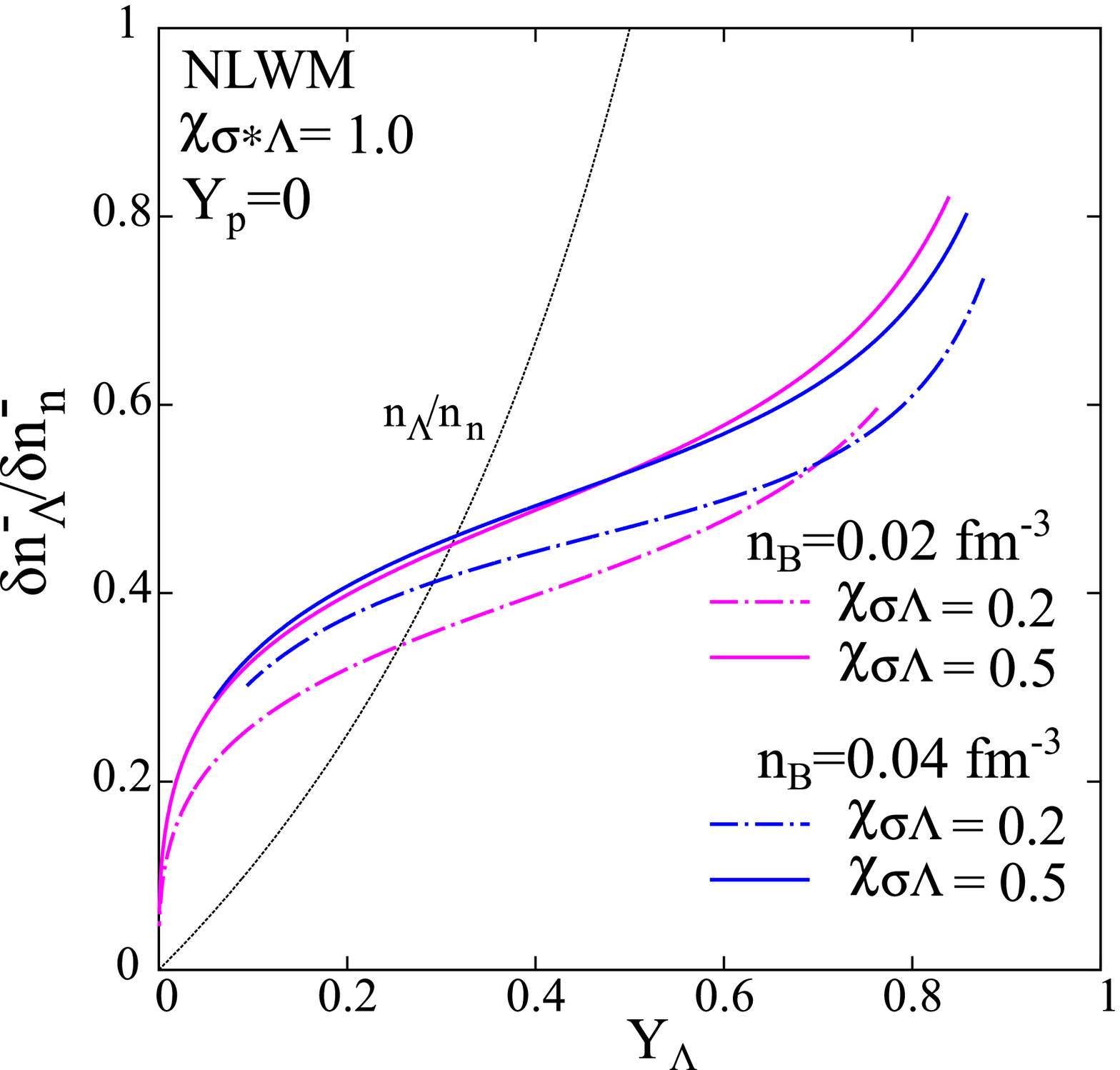}}

\caption{(Color online) Spinodals in neutron-$\Lambda$-plane with eigenvectors in NLWM. From (a) to (b) we vary $\chi_{\sigma \Lambda}$ with fixed $\chi_{\sigma^{\ast} \Lambda}=1.0$.
(c) Ratio $\delta n^{-}_{\Lambda}/\delta n^{-}_{n}$ as a function of the $Y_{\Lambda}$ for some couplings and baryon densities for $Y_{p}=0.0$.}
\label{Figure 07}
\end{figure}

\twocolumngrid

\onecolumngrid

\begin{figure}[ht!]
\centering
\subfloat[]
{\includegraphics[width=0.4\textwidth]{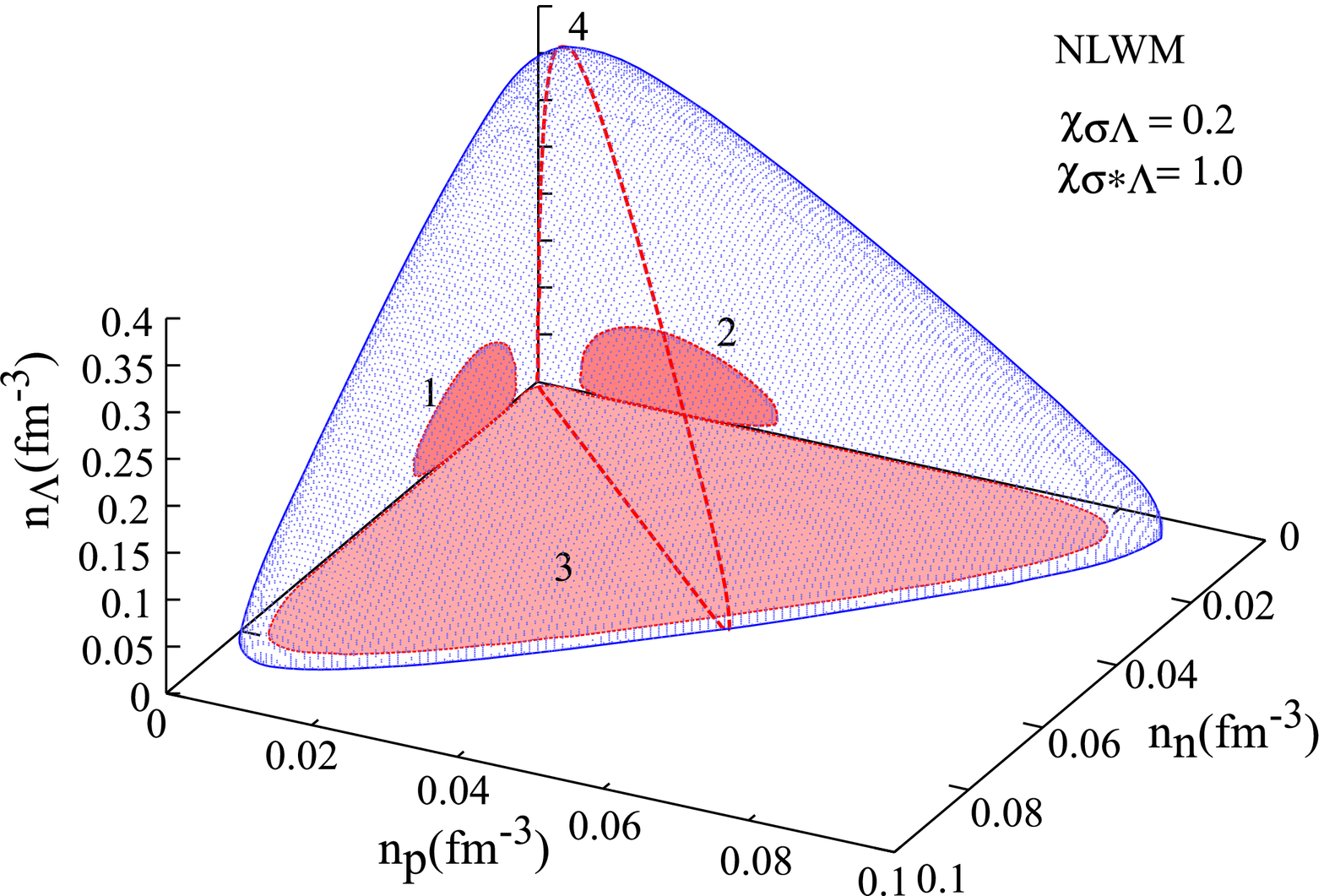}}
\subfloat[]
{\includegraphics[width=0.4\textwidth]{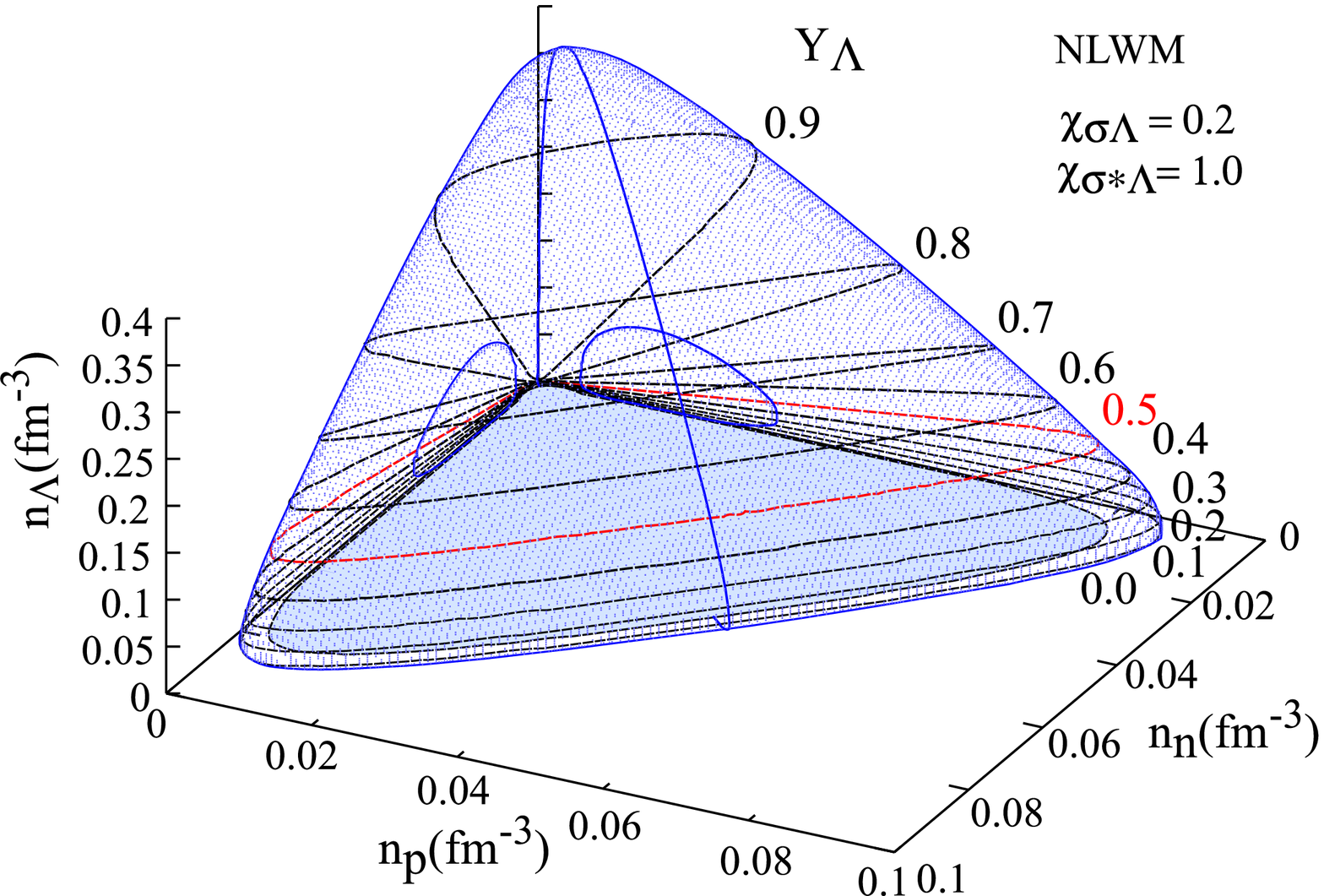}}

\caption{(Color online). Three dimensional spinodal surfaces in the NLWM for a particular choice of coupling constrained parameters. In (a) the numbers denote cuts on the surface, (1) neutron-lambda spinodal area, (2) proton-lambda spinodal area, (3) neutron-proton spinodal area
and (4) the frontier of the spinodal area when we cut the three-dimensional spinodal volume by a vertical plane passing by $n_{n}=n_{p}$. (b) show the slices when we fix $Y_{\Lambda}$ and the red shaded one is a special case.}
\label{Figure 08}
\end{figure}

\twocolumngrid
This counterintuitive result probably stems from the fact that the highest repulsion at high density is correlated to a stronger attraction at low density also in the ab-initio model. 
The behavior of the unstable eigenvector, shown in Figs.\ref{Figure 07} (a) and (b) for the two RMF parameter sets that better reproduce the ab-initio EOS, is also interesting.
We can see that it is close to the isoscalar direction $n_n+n_\Lambda$ 
as it is in the standard LG \cite{Ducoin06}. This simply means that the transition is between a dense and a diluted phase. In finite systems, the dense phase corresponds 
to an hypernucleus, and the dilute phase to a (hyper)-gas (which at $T=0$ corresponds to zero density, and wich would exist and would be in equilibrium with the 
hypernucleus at finite temperature). Fig.\ref{Figure 07} (c) shows the ratio $\delta n^{-}_{\Lambda}/\delta n^{-}_{n}$ as a function of the $Y_{\Lambda}$ for some 
couplings and baryon densities for $Y_{p}=0.0$. We can see that for very low $\Lambda$ fractions, the direction of phase separation is steeper than the constant $\Lambda$-fraction line.
This means that the dense phase is more symmetric than the dilute phase.
We also depict the line that represents $n_\Lambda/n_n$, so that it becomes visually easy to compare it with the direction of the eigenvectors.

\begin{figure}[ht!]
\centering

\subfloat[]
{\includegraphics[width=0.35\textwidth]{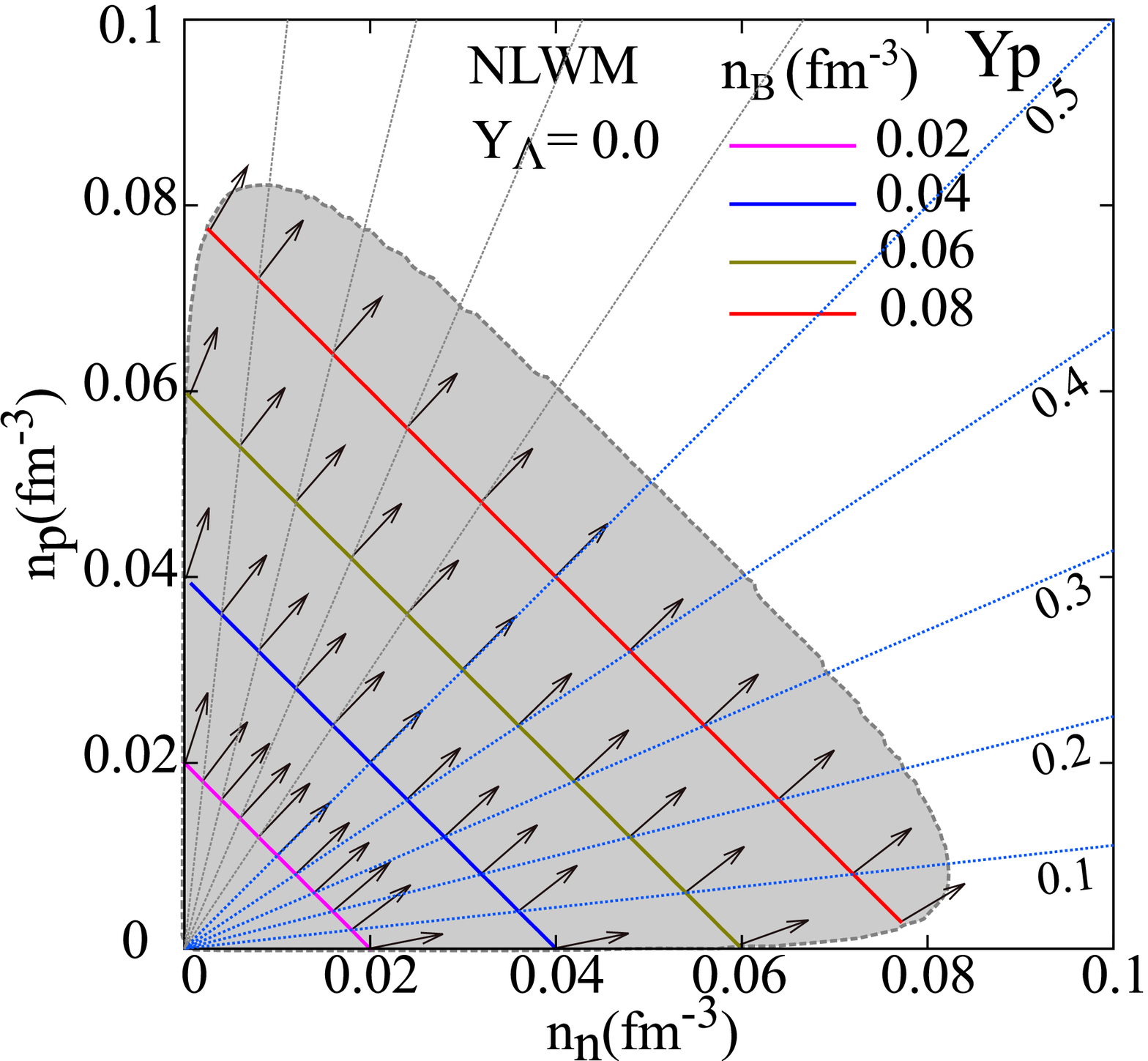}}

\subfloat[]
{\includegraphics[width=0.35\textwidth]{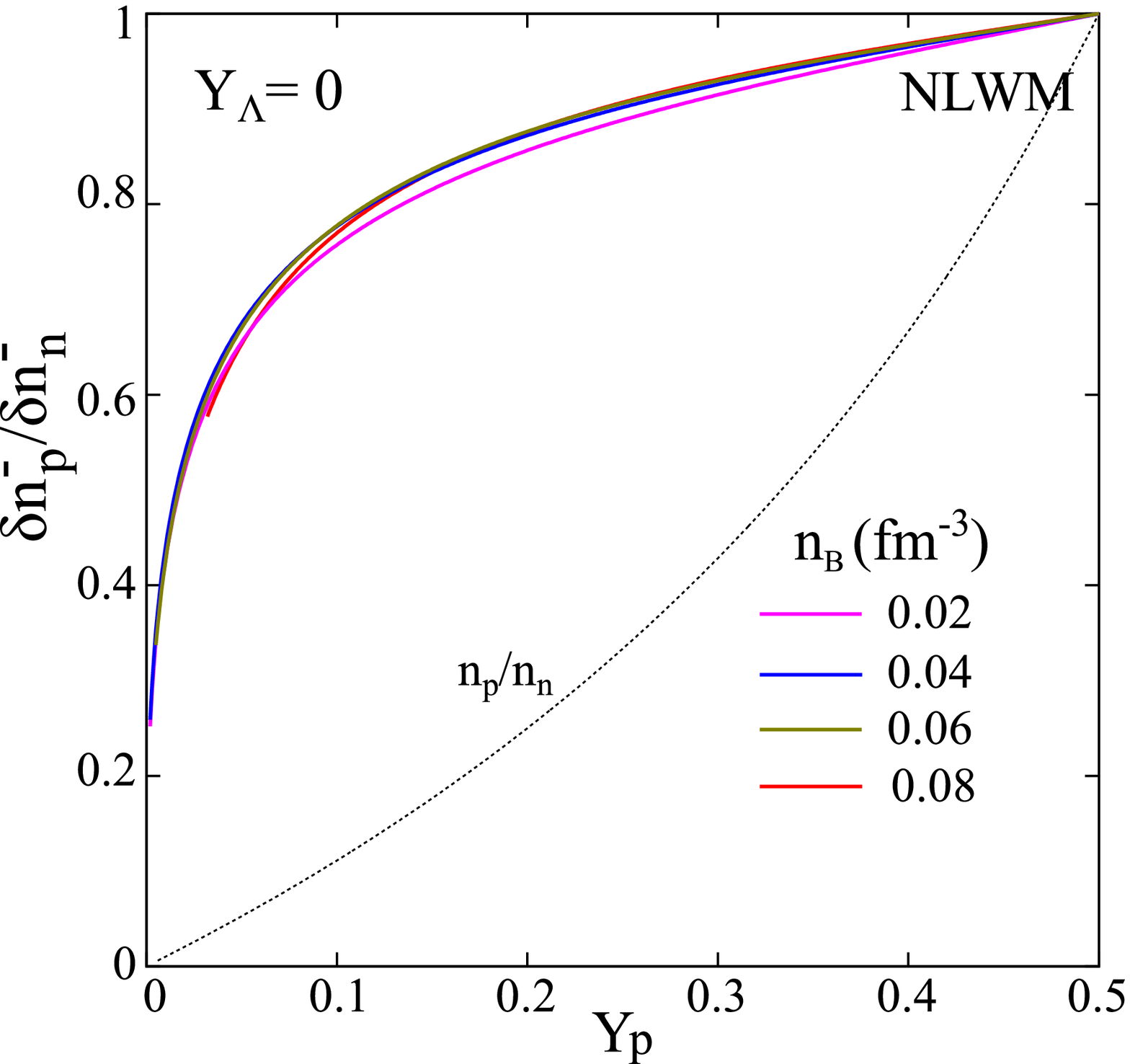}}

\caption{(Color online) (a) Spinodal for neutron-proton matter with eigenvectors in NLWM. 
(b) The ratio $\delta n^{-}_{p}/\delta n^{-}_{n}$ plotted as a function of the proton fraction for $Y_\Lambda=0$.}
\label{Figure 09}
\end{figure}

\paragraph*{}
The instability direction can be better spotted from Fig.\ref{Figure 07} (c),
which displays the unstable eigenvector as a function of the $\Lambda$ fraction. 
We can see that the unstable eigenvectors are almost independent of the baryonic density. This means that the proportion of $\Lambda$ in the dense phase 
following the spinodal decomposition is the same whatever the timescales and dynamics in the spinodal zone, and is well defined by the direction of the 
unstable eigenvectors. This proportion monotonically increase with the $\Lambda$ fraction,
but never reach the equality between neutrons and $\Lambda$. This feature is due to the mass difference between the two baryonic species, as well as to 
the reduced attraction in the $\Lambda$ channel. It is at variance with the ordinary nuclear liquid-gas which is  associated to the fractionation or 
distillation phenomenon \cite{Muller95,Ducoin06}, with the dense phase being systematically more symmetric than the dilute phase (see Fig.\ref{Figure 09}).
The optimal proportion of $\Lambda$ increases with increasing the scalar coupling, as it can be intuitively expected.

\paragraph*{}
Now we would like to see how this affects the spinodal zone calculations in the three-component system, which is more relevant for nuclear physics applications.
Fig.\ref{Figure 08} shows the three-dimensional spinodal volumes for  particular cases: $\chi_{\sigma \Lambda}=0.2$ and  $\chi_{\sigma^{\ast} \Lambda}=1.0$ 
in the NLWM. The behavior shown in both figures does not depend on the couplings used. The general pattern is always the same. 
In Fig.\ref{Figure 08} (a) the blue contour and dots are the surface
of the spinodal volume and the red shapes mean 
the slices in the orthogonal planes of this 
volume. Shape (1) represents the neutron-lambda spinodal area, (2)
proton-lambda spinodal area and (3) neutron-proton spinodal area. The
red dashed curve (4) shows the vertical plane that cuts the volume passing by $n_{n}=n_{p}$. 
Fig.\ref{Figure 08} (b) is similar to (a) but in this case the black dashed lines represent constant 
 $Y_{\Lambda}$ cuts.  $Y_{\Lambda}=0.5$ is the special value we choose
for further analysis and is highlighted in red.

\paragraph*{}
Analogue pictures for the LWM are quite similar, apart from the fact that the
size is a little bigger and no shapes (1) and (2) are present. 
Hence, in the following when we report different cuts of three-dimensional spinodal picture in RMF models, we assume that 
Fig.\ref{Figure 08} is useful to illustrate the cases LWM and NLWM. 

\paragraph*{}
A first interesting cut is at constant $\Lambda$ fraction, because it
leads to the same representation as for the usual LG phase transition, which is obtained in the limit $Y_\Lambda=0$. 
This is done in Fig.\ref{Figure 09}, which shows the spinodal region in the neutron-proton plane obtained with the NLWM model
for a large choice of coupling parameters. It is important to remark that only NLWM gives reasonable properties for symmetric matter in the absence of hyperons
and for LWM we omitted the corresponding results here.
Fig.\ref{Figure 09} (a) shows the NLWM spinodal for $Y_{\Lambda}=0$ and corresponding eigenvectors that define the region of instability analogue to the one represented by
shape (3) in Fig.\ref{Figure 08} (a). In Fig.\ref{Figure 09} (b) the ratios $\delta n^{-}_{p}/\delta n^{-}_{n}$ are plotted as a function of the proton fraction for
the same fixed baryon densities shown in Fig.\ref{Figure 09} (a).

\begin{figure*}[h!]
\centering
\subfloat[]
{\includegraphics[width=0.33\textwidth]{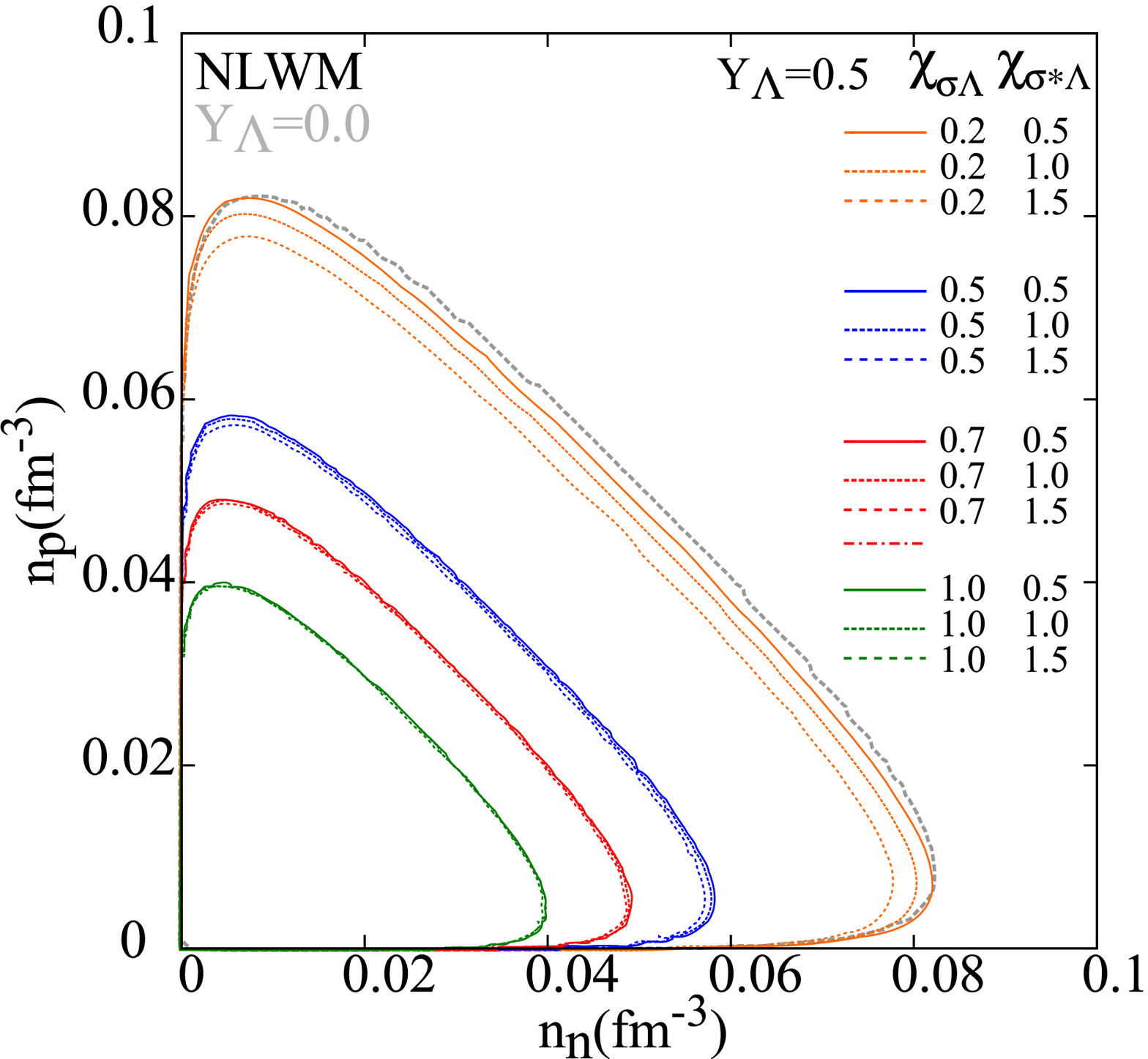}}
\subfloat[]
{\includegraphics[width=0.33\textwidth]{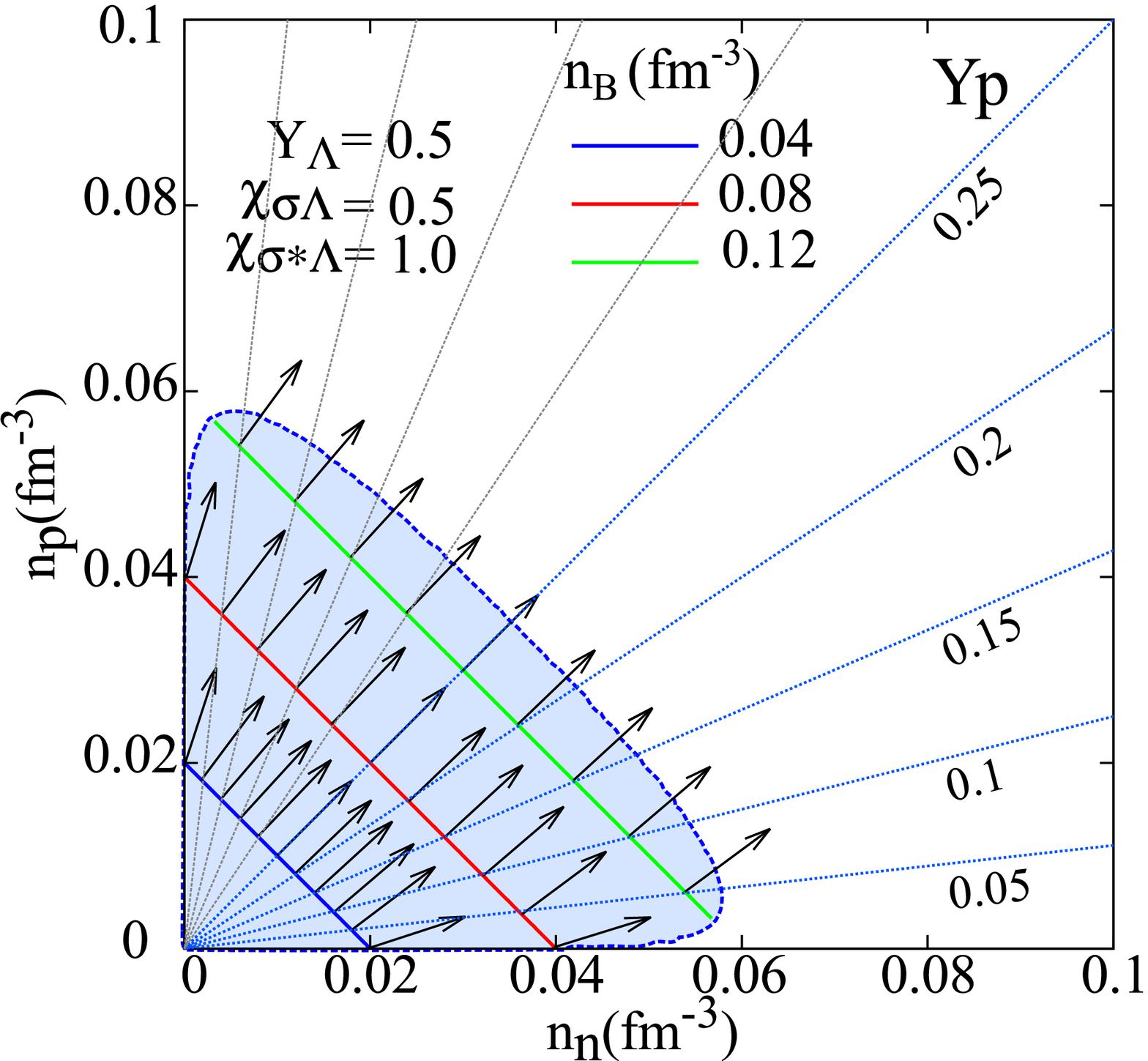}}
\subfloat[]
{\includegraphics[width=0.33\textwidth]{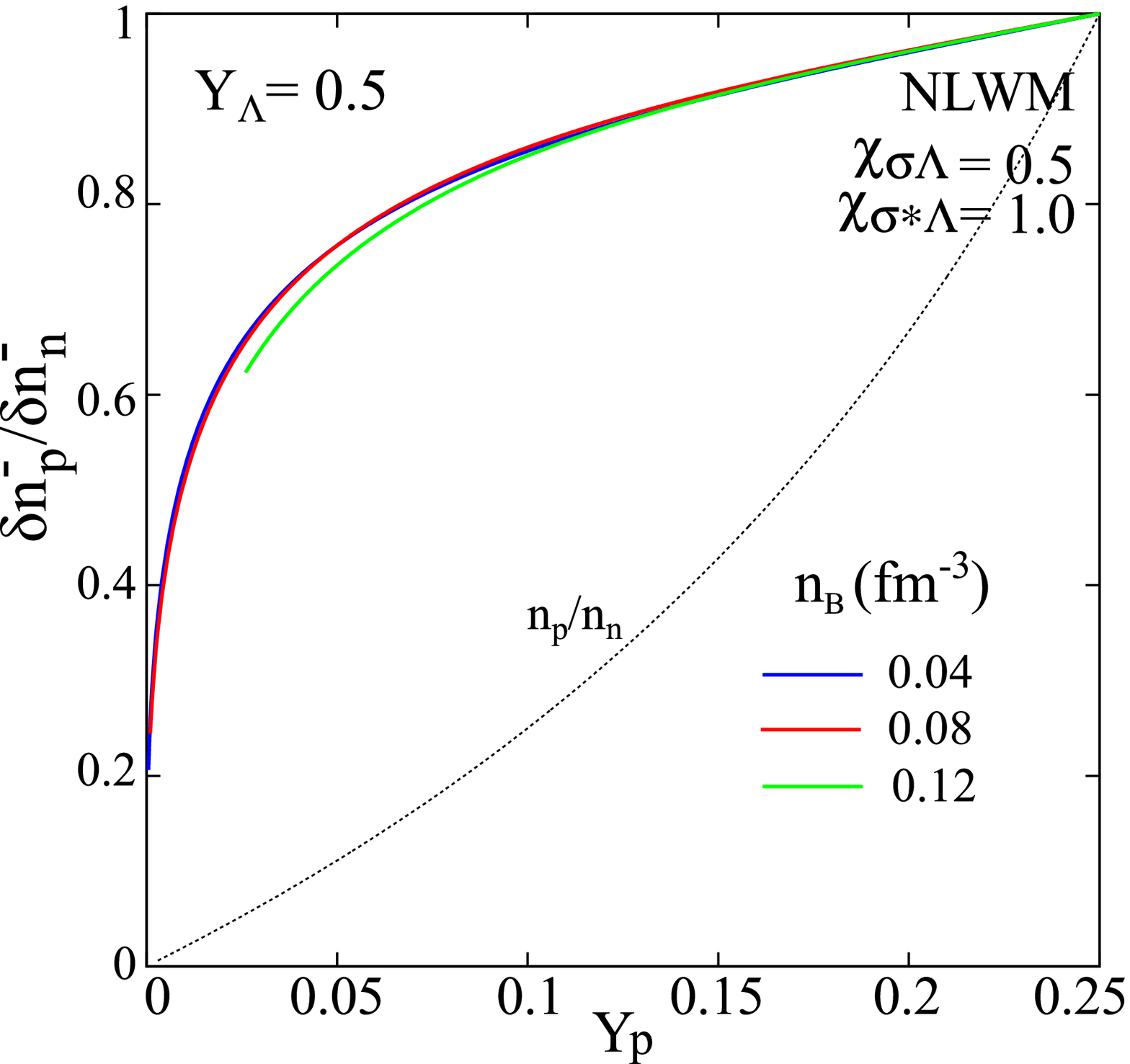}}

\caption{(Color online)  (a) Spinodals in neutron-proton plane and $Y_{\Lambda}=0.5$ in NLWM with $\chi_{\sigma \Lambda}=0.5$ and $\chi_{\sigma^{\ast} \Lambda}=1.0$. 
(b) Spinodal for neutron-proton matter with eigenvectors in NLWM.
(c) The ratio $\delta n^{-}_{p}/\delta n^{-}_{n}$ plotted as a function of the proton fraction.}
\label{Figure 10}
\end{figure*}

In Fig.\ref{Figure 10} (a) the gray curve is the frontier of the spinodal for $Y_{\Lambda}=0.0$ and the colored curves
are the spinodal frontiers obtained for $Y_{\Lambda}=0.5$ and different strange meson coupling constants in the LWM and NLWM models respectively. This colored shapes are
the projections of the spinodal curves in the neutron-proton plane for $Y_{\Lambda}=0.5$ (see Fig.\ref{Figure 08} (b)).
We recall from Section \ref{abinitio} that at low density the LWM is more realistic for the case $Y_\Lambda=0$ (left side), while the NLWM 
is in better agreement with the ab-initio model for important $\Lambda$-fractions (right side).
In any case we can see that the phase diagrams of the two models are very similar, the NLWM instability zone being only slightly narrower.

\paragraph*{}
Panels (a) and (b) of Fig.\ref{Figure 09} recall the usual characteristics of the nuclear liquid-gas phase transition \cite{Chomaz04,Ducoin06}. 
As it is well known, the instability covers a huge part of the sub-saturation region and has an essentially isoscalar character. 
The unstable eigenvectors point towards a direction which is intermediate between the isoscalar direction (observed only for symmetric matter $n_n=n_p$), 
and the direction of constant isospin. As a consequence, the dense phase is systematically more symmetric than the dilute phase. 
Indeed, at zero temperature the dilute phase is a pure gas of neutrons (protons) if the system is neutron (proton) rich \cite{Ducoin06}.
From panels (a) and (c) of Fig.\ref{Figure 10} we additionally learn that
the LG instability is clearly preserved by the addition of strangeness.
However, the transition is quenched for strongly coupled hyperons. Indeed,
we can clearly see that when $\chi_{\sigma \Lambda}$ increases
the spinodal area decreases. 
Considering that the most realistic value lays around
$\chi_{\sigma \Lambda}\approx 0.2-0.5$, this quenching is small.
On the other side, when $\chi_{\sigma^{\ast} \Lambda}$ increases, the modification on the spinodal is very small. 
This is expected, since 
the strange mesons are only coupled to strange baryons and are therefore
expected to affect essentially the $\Lambda$ density, 
which is not represented here.
Due to the weak effect of $\chi_{\sigma^{\ast} \Lambda}$ in the spinodal
frontier we select the value $\chi_{\sigma^{\ast} \Lambda}=1.0$ 
to study the eigenvectors in the neutron-proton plane displayed in the
next figures.
For the NLWM spinodal area shown in Fig.\ref{Figure 10} (b) and the 
vectors represent the projection of the unstable eigenvectors on the neutron-proton plane.
In Fig.\ref{Figure 10} (c) the ratio $\delta n^{-}_{p}/\delta n^{-}_{n}$ are plotted as a function of the proton fraction.  
No difference can be seen with respect to the normal LG: whatever the percentage of $\Lambda$'s,
the neutron-proton composition of the dense phase (i.e. the hypernucleus)
is unmodified, even if the density is reduced. 

\onecolumngrid

\begin{figure}[ht!]
\centering
\subfloat[]
{\includegraphics[width=0.45\textwidth]{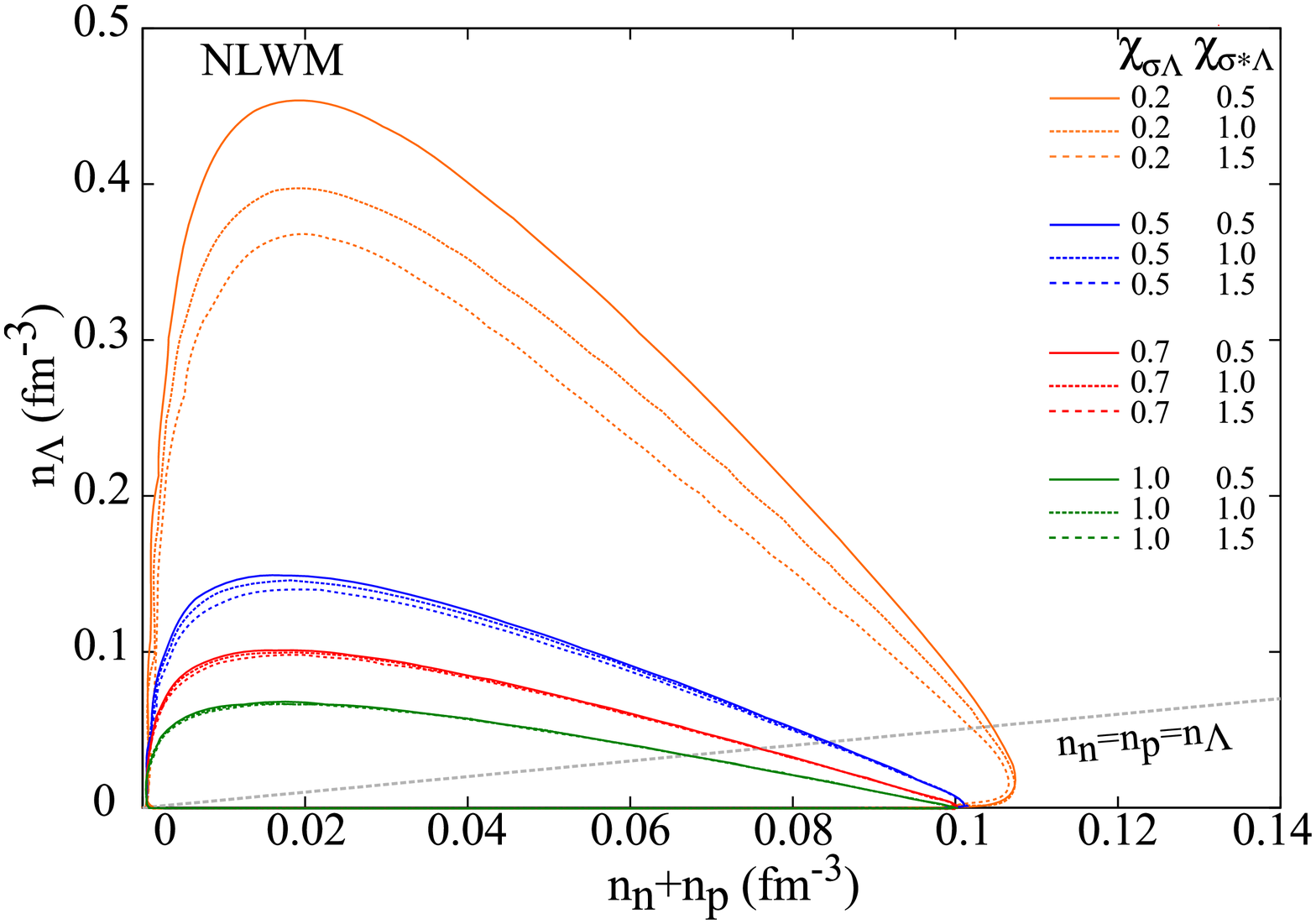}}

\caption{(Color online) Spinodal in the nucleon-$\Lambda$-plane (keeping $n_{n}=n_{p}$)  (a) Colors contours are sliced shapes from 3d spinodal in densities space and varying $\chi_{\sigma \Lambda}$
and $\chi_{\sigma^{\ast} \Lambda}$ in NLWM. The gray dotted line represents the $n_N=n_\Lambda$ line.}
\label{Figure 11}
\end{figure}

\twocolumngrid
\paragraph*{}
This finding might seem in contradiction with recent studies in multiply strange hypernuclei \cite{Ikram14,Sang14,Khan15}, where it is seen 
that the driplines are modified by the $\Lambda$-fraction. However these modifications are essentially due to shell and Coulomb effects, which are not accounted for in this infinite matter calculation.
If we change our perspective from the neutron-proton plane
to imagine the general three-dimensional spinodal \textit{locus} and
instead of fixing $Y_{\Lambda}$ as before, 
we fix the symmetric matter condition $n_{N}=2n_{n}=2n_{p}$, 
the resulting plane slice crossing this three-dimensional volume 
is similar to the curve denoted by number (4) in Fig.\ref{Figure 08} (a). The related spinodal areas for the NLWM and many choices of the coupling parameters
are shown in Fig.\ref{Figure 11}. 

\paragraph*{}

\begin{figure*}[ht!]
\centering
\subfloat[]
{\includegraphics[width=0.32\textwidth]{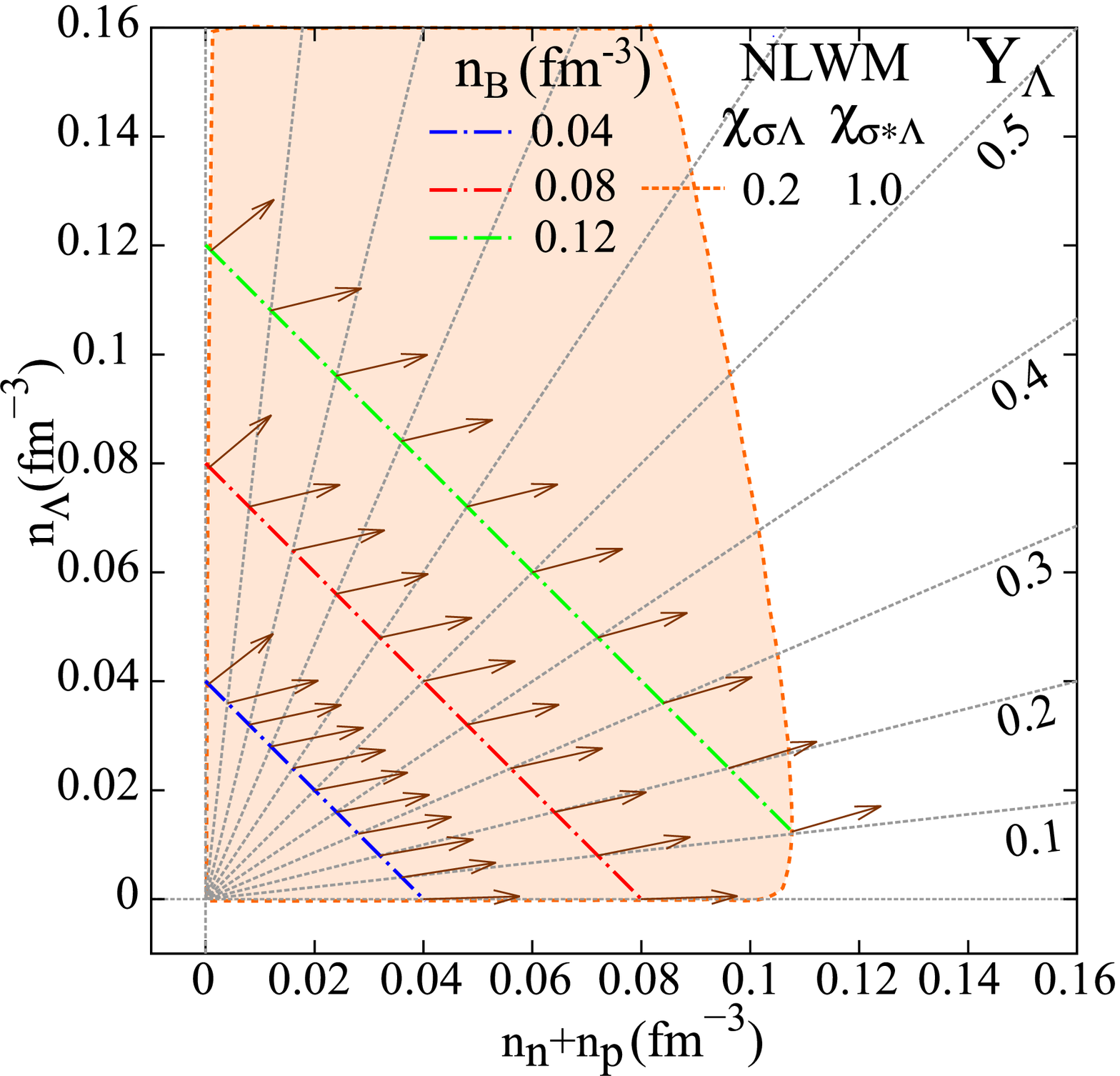}}
\subfloat[]
{\includegraphics[width=0.32\textwidth]{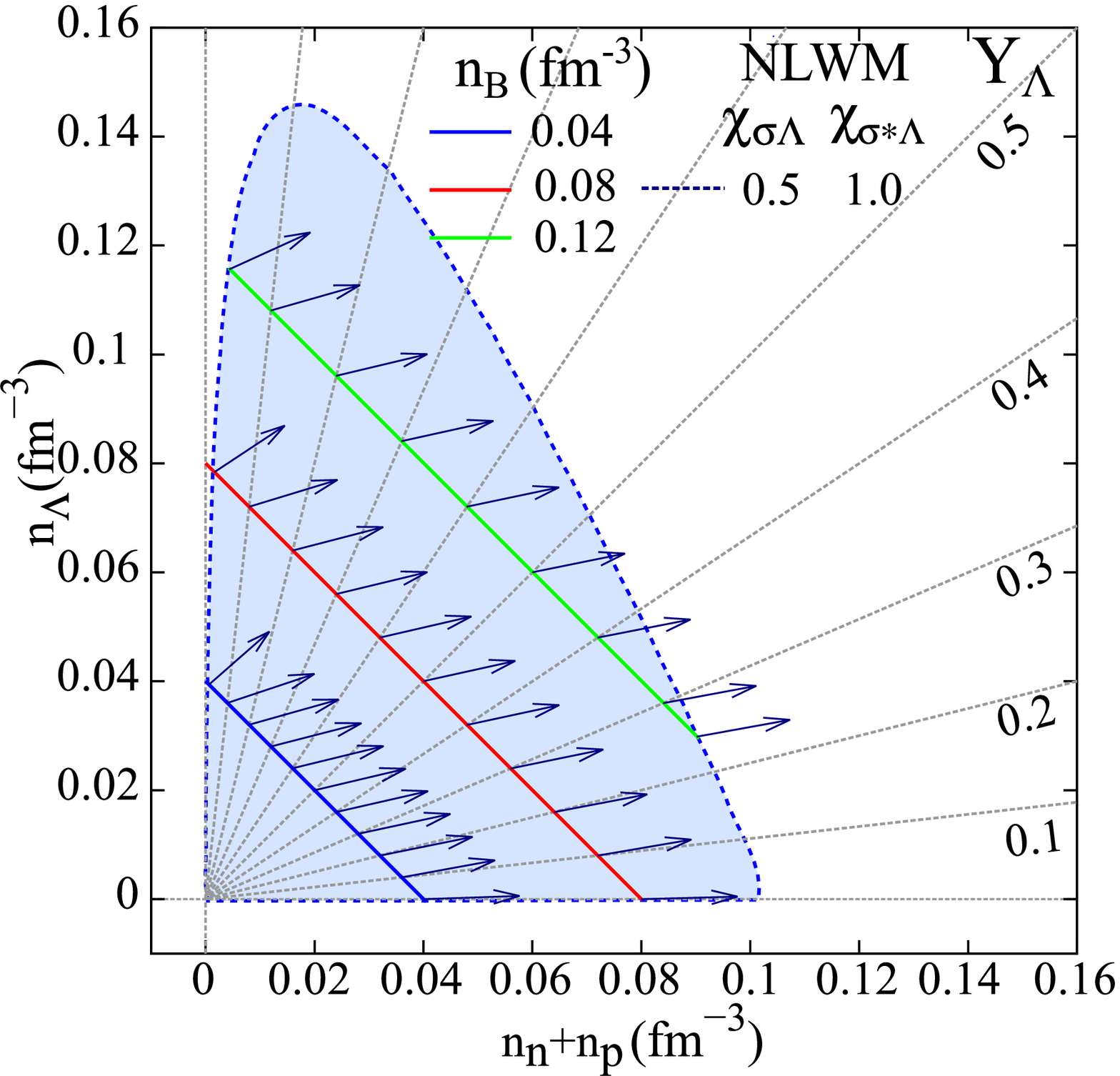}}
\subfloat[]
{\includegraphics[width=0.32\textwidth]{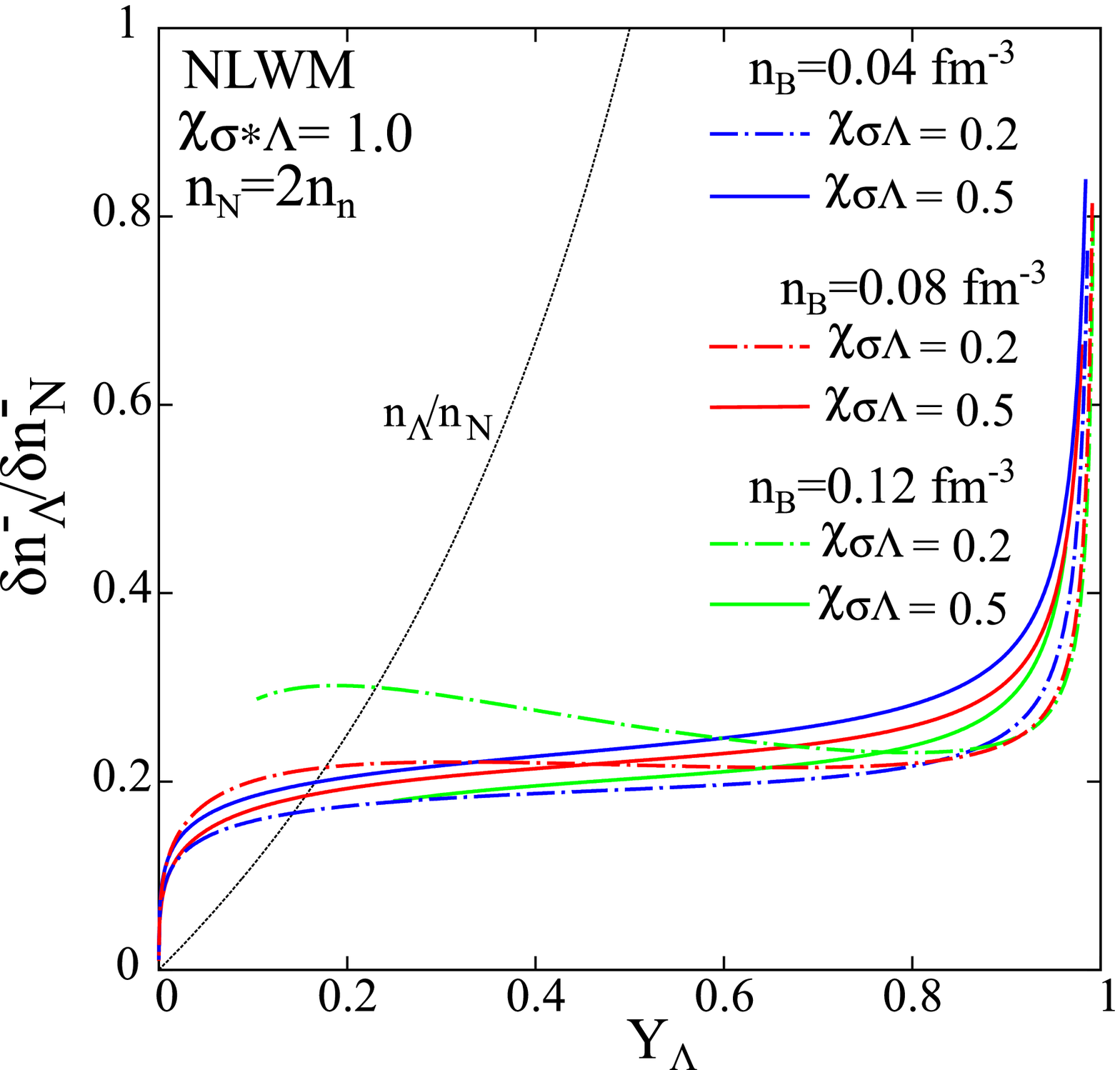}}

\caption{(Color online) Spinodals in nucleon-$\Lambda$-plane ($n_{n}=n_{p}$) with eigenvectors in NLWM. 
From (a) to (b) we vary $\chi_{\sigma \Lambda}$ with fixed $\chi_{\sigma^{\ast} \Lambda}=1.0$. (c) Ratio $\delta n^{-}_{\Lambda}/\delta n^{-}_{N}$.} 
\label{Figure 12}
\end{figure*}

The comparison to the ab-initio model of Section \ref{abinitio}
suggests that the most realistic phase diagram should be between the ones corresponding to 
$\chi_{\sigma\Lambda}=0.2, 0.5$, which gives an energy functional intermediate
between the two AFDMC parametrizations of three-body forces.
We can see that the coupling to the strange meson $\chi_{\sigma^{\ast} \Lambda}$ has a bigger effect in this plane as expected.
Still, its influence on the spinodal is small. This means that the wide uncertainty on the strange mesons has a negligible influence 
on the phase transition. The biggest uncertainty concerns the extension of the spinodal zone along the $n_\Lambda$ axis. It is however important to stress that 
this situation $n_N<n_\Lambda<n_0$ does not correspond to any known physical system.
Every shape shown touches the horizontal axis when $Y_{p}=0.5$, as it should be considering that 
the np system is bound. At this point we can report to Fig.\ref{Figure 03}, to see that our restriction of $\chi_{\sigma \Lambda}$ does not affect much the
spinodal zone analysis because when we increase $\chi_{\sigma \Lambda}$ up to $1.3$ the spinodal zone tends to become flatter in the $\Lambda$-density direction. 
Even if the calculation might be not realistic for very high $\Lambda$-fraction, we can conclude that the LG phase transition is still present in multistrange systems.
\paragraph*{}
Finally, Figs.\ref{Figure 12} (a) and (b) show the projections of the unstable eigenvectors in the $\Lambda$-nucleon plane.
We can see that a non negligible component of the order parameter lies along the $n_\Lambda$ direction, meaning that the $\Lambda$-density is an order parameter of the phase transition, 
or in other words that the dense phase is also the phase with the higher strangeness content. 
These eigenvectors are almost parallel to each other, and considerably deviate
with respect to the direction of the constant $\Lambda$-fraction lines as seen in Fig.\ref{Figure 12} (c).
Interesting enough, the instability points towards an ``optimal" composition $n_\Lambda\approx 0.2 n_N$ for the dense phase,
whatever the baryonic density, coupling constants and $\Lambda$ fraction. Only for very small and very high $\Lambda$ fraction 
a deviation is observed. This is expected because by construction the instability must tend towards the non-strange direction 
in the absence of strangeness. It will be very interesting to verify if such an optimal composition is obtained in calculations of multiple-strange hypernuclei.
As in the case of the simpler $n-\Lambda$ system, the fact that the instability 
always points towards $\Lambda$ poor systems is
at variance with the distillation phenomenon, characteristic of the LG phase transition with more than one component \cite{Muller95,Ducoin06}, where 
the direction of phase separation tends to equal composition. This symmetry breaking between nucleons and $\Lambda$'s comes from the difference in the bare mass 
of the particles and the less attractive couplings. 
Still, for very low $\Lambda$ fractions, the direction of phase separation is steeper than the constant $\Lambda$-fraction line.
This means that the dense phase is more symmetric than the dilute phase. 
This thermodynamic finding is compatible with the observation
in Ref.\cite{Mallik15} that the  $\Lambda$'s produced in heavy-ion collisions should stick to the clusters (i.e., the dense phase) rather than being emitted as free 
particles (i.e., the gas).

\onecolumngrid

\twocolumngrid

\section{Summary and Conclusions} \label{summary}

\paragraph*{}
We have investigated the thermodynamic phase diagram at subsaturation density, for baryonic matter including neutrons, protons and $\Lambda$ hyperons, within a
RMF approach. For the nucleonic EOS, we have considered the GM1
parametrization of NLWM, together with the simpler LWM. 
Strange mesons were included to allow a wide exploration of the possible phenomenology for the (still largely unknown) hyperon-nucleon and hyperon-hyperon couplings, 
with minimal requirements on the potential depths extracted from hypernuclear data. 
Imposing these requirements leads to a strong linear correlation between 
the attractive and the repulsive couplings, both for the normal and the strange mesons.
These constraints leave us with a two-dimensional parameter space, which we have varied widely in order to pin down generic features of the phase diagram.
\paragraph*{}
Our main focus was the understanding  of the instabilities in the
hypernuclear matter, and specifically the influence of $\Lambda$'s in the well known Liquid-Gas phase transition of nuclear matter. 
The existence of an instability as a signature of
a first order phase transition 
was identified by analyzing the curvature of the thermodynamic
potential with respect to the nucleonic and strange densities. 
In all our studies one and only one negative eigenvalue has
been found, showing that the phase transition still exists in the presence of strangeness and is still of LG type, even if its extension in the density space shrinks with increasing strangeness.
The negative eigenvalue corresponds to
the direction in density space, in which 
density fluctuations get spontaneously 
and exponentially amplified in order to achieve phase separation. 
This eigenvalue is seen to systematically have a non-negligible component in the direction of the strange density. This means that strangeness can be viewed as an order parameter of the transition. 
\paragraph*{}
Less expected is the fact that the instability direction systematically points to a fixed
proportion of $\Lambda$'s in the dense phase, at variance with 
the phenomenon of distillation typical of binary mixtures.
This proportion being of the order of 30\% in the models we considered, this means
that in a dilute system with a small contribution of $\Lambda$'s, these latter will preferentially belong to the dense clusterized phase.
These conclusions are general and appear largely model independent. 
On the contrary, the specific shape of the phase diagram would obviously depend
on the choice of the free $\chi_{\sigma^{\ast}\Lambda}$ and $\chi_{\phi
 \Lambda}$  couplings.
Some hints on a more quantitative estimation of the thermodynamics 
were obtained from the analysis  of the simpler n$\Lambda$ phase diagram
extracted from the ab-initio AFDMC calculation of Ref.\cite{abinitio00}.
 The characteristics of the phase transition are confirmed in the ab-initio model,
even if the phase diagram extension depends on the three-body force model
in an important way.
\paragraph*{}
%
  
The comparison of the RMF with the AFDMC also reveals
some limitations of the phenomenological model at low density.
Indeed the popular GM1 model is shown to compare
very poorely to the ab-initio calculation of pure neutron matter
even at the low densities considered in the present study.
Unexpectedly, the simpler LWM is in very good agreement
with the ab-initio predictions at low density. Concerning the
n$\Lambda$ mixture, the energy functional is within the theoretical
error bars if $0.2\leq\chi_{\sigma\Lambda}<0.5$..
Other parameterizations  could change the
quantitative results that we have presented in this paper.
In particular a recent work \cite{odilon} shows
that the scalar-isovector $\delta$ meson also plays an important
role in satisfying both nuclear bulk and stellar properties
constraints. The use of another parameter set and/or
the inclusion of this new degree of freedom
requires a complete calculation from the very beginning because
the nucleon-lambda potential, Eq.(\ref{Uln_contraint_densities}), has to be readjusted.
However, the qualitative results will be certainly
similar, since the whole 3D parameter space associated
to strangeness was spanned.
As a perspective for future
work, it will be very interesting to analyze the instability behavior
of a density dependent coupling RMF model, directly
fitted to the ab-initio calculation.

\begin{acknowledgments}
This work was partially supported by CAPES/COFECUB project 853/15, CNPq under grants 300602/2009-0 and 470366/2012-5.
\end{acknowledgments}

\end{document}